\documentclass[conference]{IEEEtran}

\let\labelindent\relax
\usepackage{enumitem}
\usepackage{tabularx}
\usepackage{graphicx}
\usepackage{mdframed}
\usepackage{xcolor}
\usepackage{tcolorbox}
\usepackage{makecell}
\usepackage{listings}
\usepackage{array}     
\usepackage{afterpage}
\usepackage{float}
\usepackage{caption}   
\usepackage{subcaption}
\usepackage{tikz}  
\usepackage{lipsum}  
\usepackage{booktabs}
\usepackage{multirow}
\usepackage{colortbl}
\usepackage{xspace}
\usepackage{pifont}
\usepackage{threeparttable}
\usepackage{adjustbox}
\usepackage{listings}
\usepackage{tikz}
\usepackage{pgf}
\usepackage{tcolorbox}
\usepackage{multicol}
\usepackage{hyperref}

\ifCLASSOPTIONcompsoc
  \usepackage[nocompress]{cite}
\else
  \usepackage{cite}
\fi

\ifCLASSINFOpdf
\else
\fi

\newcommand{\sysname}  {\textsc{Vulture}\xspace}
\newcommand{\tplname}
{\textsc{TPLFilter}\xspace}

\newcommand{\tool}[1]{{\small{\textsf{{#1}}}}}
\newcommand{\func}[1]{{\texttt{{#1}}}}

\definecolor{myblue}{RGB}{70,130,180}
\definecolor{darkred}{RGB}{139, 0, 0}

\definecolor{commentgreen}{RGB}{2,112,10}
\definecolor{eminence}{RGB}{108,48,130}
\definecolor{weborange}{RGB}{255,165,0}
\definecolor{frenchplum}{RGB}{129,20,83}

\begin{document}

\title{Enhancing Security in Third-Party Library Reuse - Comprehensive Detection of 1-day Vulnerability through Code Patch Analysis}

\author{
    \IEEEauthorblockN{Shangzhi Xu\IEEEauthorrefmark{4}, 
                      Jialiang Dong\IEEEauthorrefmark{4}, 
                      Weiting Cai\IEEEauthorrefmark{2}, 
                      Juanru Li\IEEEauthorrefmark{3}, 
                      Arash Shaghaghi\IEEEauthorrefmark{4}, 
                      Nan Sun\IEEEauthorrefmark{4}, 
                      Siqi Ma\IEEEauthorrefmark{4}\textsuperscript{*}}
    \IEEEauthorblockA{\IEEEauthorrefmark{4}The University of New South Wales, Sydney, Australia\\
                      Emails: \{shangzhi.xu, jialiang.dong, a.shaghaghi, nan.sun, siqi.ma\}@unsw.edu.au}
    \IEEEauthorblockA{\IEEEauthorrefmark{2}Delft University of Technology, Delft, Netherlands\\
                      Email: weitingcai2020@gmail.com}
    \IEEEauthorblockA{\IEEEauthorrefmark{3}Feiyu Technology International Company Ltd\\
                      Email: romangol\_t@hotmail.com}
}

\thanks{\textsuperscript{*}Corresponding author.}

\IEEEoverridecommandlockouts
\makeatletter\def\@IEEEpubidpullup{6.5\baselineskip}\makeatother
\IEEEpubid{\parbox{\columnwidth}{
    Network and Distributed System Security (NDSS) Symposium 2025\\
    24 - 28 February 2025, San Diego, CA, USA\\
    ISBN 979-8-9894372-8-3\\
    https://dx.doi.org/10.14722/ndss.2025.240576\\
    www.ndss-symposium.org
}
\hspace{\columnsep}\makebox[\columnwidth]{}}

\maketitle

\begin{abstract}
Nowadays, 
    software development progresses rapidly to incorporate new features. 
To facilitate such growth and provide convenience for developers when creating and updating software, 
    reusing open-source software (i.e., third-party library reuses) has become one of the most effective and efficient methods. 
Unfortunately, 
    the practice of reusing third-party libraries (TPLs) can also introduce vulnerabilities (known as 1-day vulnerabilities) because of the low maintenance of TPLs, resulting in many vulnerable versions remaining in use.
If the software incorporating these TPLs fails to detect the introduced vulnerabilities and leads to delayed updates,
    it will exacerbate the security risks.  
However, 
    the complicated code dependencies and flexibility of TPL reuses make the detection of 1-day vulnerability a challenging task. 
To support developers in securely reusing TPLs during software development,
    we design and implement \sysname, 
    an effective and efficient detection tool, 
    aiming at identifying 1-day vulnerabilities that arise from the reuse of vulnerable TPLs.  
It first executes a database creation method, \tplname, 
    which leverages the Large Language Model (LLM) to automatically build a unique database for the targeted platform.
Instead of relying on code-level similarity comparison, 
    \sysname employs hashing-based comparison to explore the dependencies among the collected TPLs and identify the similarities between the TPLs and the target projects. 
Recognizing that developers have the flexibility to reuse TPLs exactly or in a custom manner, 
    \sysname separately conducts version-based comparison and chunk-based analysis to capture fine-grained semantic features at the function levels.  

We applied \sysname to 10 real-world projects to assess its effectiveness and efficiency in detecting 1-day vulnerabilities. 
\sysname successfully identified 175 vulnerabilities from 178 reused TPLs. 

\end{abstract}

\IEEEpeerreviewmaketitle

\section{Introduction}
\label{sec:introduction}

As software evolves various innovative functionalities nowadays (e.g., AI-based classification, unmanned operations), 
    it makes software increasingly complicated and challenging to develop and maintain because of intricate dependencies among its massive functions.
Open-source software (OSS) reuse enables developers to integrate the functionalities faster,
    which facilitates more efficient development and code maintenance. 
Simultaneously, 
    reusing OSS supports flexible development of features because developers can either deploy OSS as third-party libraries (TPLs) exactly or in a custom manner~\cite{v1scan}.  

However, 
    this convenience also makes it more likely for developers to unintentionally introduce vulnerabilities, known as 1-day vulnerabilities, through TPL reuse~\cite{endorlabs2023}~\cite{tang2022towards}.
Such security issues are commonly brought because 1) some TPLs may no longer be actively maintained anymore, leaving their functionalities incomplete or vulnerable to security risks, especially when vulnerabilities are exploited;
    2) the decentralized nature of OSSes makes it difficult to keep track of all code changes made by contributors, 
    complicating the enforcement of security reviews and practices when reusing the TPLs;
    3) TPLs are not always developed by following the best practices, 
    which can make them more vulnerable. 
For instance, 
    MOVEit Transfer, a file transfer project, experienced a security breach due to SQL injection in June 2023, 
    which puts all dependent software at risk of unauthorized access; thus the dependent software is required to install the patched version to safeguard their data promptly~\cite{progress2023}.

To explore 1-day vulnerabilities brought by TPL reuses, 
    some studies~\cite{redebug}~\cite{vuddy}~\cite{Centris} only analyze the reuses without any modifications (i.e. exact reuse) by employing similarity comparison, 
    which identifies whether the functions from TPLs are invoked in the target program. 
Unfortunately, 
    exact reuse is only a small portion of TPL reuse because of its functional restrictions during software development. 

Such flexibility of custom TPL reuse enhances functionality customization, 
    yet it simultaneously introduces challenges in recognizing TPL reuse and detecting 1-day vulnerabilities caused by TPL reuse.
Although some existing tools, such as \tool{V1SCAN}~\cite{v1scan} and \tool{MVP}~\cite{mvp}, claimed that they are capable of analyzing custom TPL deployment, 
    these tools can only handle simple custom reuses with minor modifications,
    which leaves the problem far away from being fully resolved. 
Making 1-day vulnerability detection more complicated, 
    developers may fix vulnerable TPL functions in a custom way, 
    rather than using official updates or upgrading the entire TPLs~\cite{OSSPolice}.

To bridge the above gaps, 
    we propose \sysname, a novel vulnerability detection tool, to explore the potential 1-day vulnerabilities brought by TPL reuses \emph{effectively} and \emph{efficiently}. 
As TPLs designed for different target platforms vary significantly~\cite{zeng2024survey}, 
    \sysname first conducts a mutually promoted approach, \tplname, to heuristically refine existing OSSes by selecting the TPLs appropriate for the specific target platform(s).
It constructs a multifaceted database that includes all the vulnerable and patched TPLs through LLM-based commit slicing. 
Instead of retaining all information (e.g., project code, descriptions, fix commits) of each TPL, 
    \tplname condenses each TPL version into several independent functions, 
    and then converts the functions into compact and numerical tuples using the LSH algorithm~\cite{jafari2021survey}.
According to the TPL database generated by \tplname, \sysname employs a similarity-based method to detect TPL reuse within the target program, identifying both the reused TPLs and their versions. 

Subsequently, 
    it undertakes a dual TPL analysis to meticulously examine exact and custom code reuse.
In particular, 
    \sysname first conducts version-based analysis to explore exact TPL reuses by identifying the vulnerable TPL versions from the target project.
Then, 
    it employs fine-grained reformatting through code tokenization and chunk-based analysis to identify custom TPL reuse.
To preserve semantic and contextual information,
    \sysname generates each chunk by conducting intra-procedural analysis to collect information on variables (i.e., values, operations, and relative positions) involved in code modifications during the transition from the vulnerable code to the patched code.
By comparing the chunks with the target program, 
    \sysname reports whether any 1-day vulnerabilities exist and pinpoints their exact locations.


To assess the effectiveness of \tplname and \sysname, 
    we created a benchmark by manually analyzing 68 real-world projects,
    labeling 200 TPL reuses, 
    and identifying 200 vulnerable reused functions from those reuses.
By integrating the database created by \tplname, 
    \sysname successfully identified 184 vulnerable reused functions, achieving an F1 @ 95.8\% while the state-of-the-art tool, \tool{V1SCAN} only detected 100 vulnerable reused functions, achieving an F1 @ 66.7\%.
\sysname not only outperforms the state-of-the-art academic tool but also surpasses the commercial tool.
We applied \sysname and a commercial detection tool \tool{SNYK} \cite{Snyk_Vulnerability_Database} to analyze TPL reuses within 10 real-world projects on a large scale.  
\sysname outperformed \tool{SNYK} by identifying 175 vulnerabilities from 178 TPL reuses, while \tool{SNYK} identified 111 vulnerabilities and \tool{V1SCAN} only identified 13.
Our findings indicate that custom adaptations of TPLs are widespread, representing about 55\% of all reuses. Additionally, the majority of 1-day vulnerabilities were due to the reuse of outdated TPLs, a problem often compounded by poor maintenance and the lack of thorough fix guidance from public security advisories.


\noindent
\textbf{Contribution:}

\begin{itemize}

\item \emph{A novel approach to automatically build an extendable database tailored for TPL analysis on a designated platform.}

We design a mutually promoted approach, \tplname,
    that integrates keyword searching with a LLM to explore the commonly used TPLs, 
    along with the associated vulnerabilities and patch code. 
All vulnerabilities and patch information are then gathered to create a TPL vulnerability database.

\item 
\emph{An effective 1-day vulnerability detection tool to discover the reused TPLs and any potential vulnerabilities that have been exploited.}

We design a 1-day vulnerability detection tool, \sysname, 
    that employs locality-sensitive hashing (LSH) comparisons and a dual intra-procedural analysis to exploit target programs semantically and syntactically. 
It can distinguish the vulnerable TPLs by recognizing both the official patches provided by TPL owners and the custom patches created by developers.



\item 
\emph{A comprehensive evaluation across \tplname and \sysname.} 
We evaluate both \tplname and \sysname by comparing them with state-of-the-art tools on real-world projects. 
\sysname discovered 175 vulnerabilities from 10 real-world IoT projects and provided 154 patch commits.
    
\end{itemize}

\vspace{0.05cm}
\noindent
\textbf{Availability.}
The source code of \sysname and the experiment datasets are available at
\href{https://anonymous.4open.science/r/Vulture-17BC}{https://anonymous.4open.science/r/Vulture-17BC}.

\section{Background}
\label{sec:background}


\subsection{Third-party Library Reuse}
Developers extensively employ TPLs to cater to the diverse requirements of different users. 
For example, 
    multiple MQTT client libraries are designed to simplify the deployment and implementation for connecting remote devices such as IoT devices, while \tool{zlib} is frequently employed for managing file compression and decompression.

To recognize used TPLs, 
    state-of-the-art detectors usually operate across three phases. 
First,
    a TPL database containing whitelists of known libraries is built. 
These whitelists are typically generated through manual analysis and require regular updates. 
Given the database, 
    detectors then collect the representative features/signatures (e.g., invoked functions~\cite{libD, Centris, OSSFP}, keyword tokens~\cite{SourcererCC, CCAligner}, function dependencies~\cite{TPLite, BCFinder, LibAM, LibvDiff, Gemini}) of the OSS utilized as the target for TPL reuse detection (i.e. target program).
To enhance the effectiveness of TPL reuse detection, 
    some detectors optimize the process by eliminating redundant functions and statements~\cite{Centris} or defining the significance level of each function~\cite{OSSFP, BCFinder}. 
The similarity score between the collected representative features/signatures and the libraries stored in the database is further calculated. 
Reuse is confirmed if the similarity score exceeds a pre-defined threshold. 
The major challenge in detecting reused TPLs is \emph{the inconsistency of reusing TPLs within different targeted programs}. 

\vspace{0.05cm}
\noindent
\textbf{TPL database construction.}
\label{tpl_construction}
Without a universal standard, 
    developers across different platforms (e.g., IoT firmware, mobile operating systems, Open-source software) tailor varied infrastructures to meet specific objectives.
Hence,
    TPL providers sometimes offer specialized versions of TPLs to better adapt to specific platform(s). 
For instance, 
  \tool{CocoaMQTT} ~\cite{CocoaMQTT}, \tool{Paho Android Service}~\cite{mqttAndroid} and \tool{Paho MQTT C/C++ client}~\cite{mqttembed} are MQTT client libraries for iOS, Android, and embedded platforms, respectively. 
Hence, 
    the TPL database used for reuse matching must be:
\begin{itemize}
\item \textbf{Comprehensive:} 
TPLs that are commonly invoked within the targeted platform must be included.

\item 
\textbf{Specific:}
Different platforms may have unique libraries that cannot be used by the other platforms. 
To avoid false alarms, libraries specific to other platforms, unnecessary for the targeted platform, or projects not used as libraries must be excluded during the detection process. 
    
\item 
\textbf{Maintainable:}
TPLs are often developed to accelerate the development cycle of open-source software and are regularly updated to fix bugs and integrate new features. 
Hence, 
    the libraries included in the database should also be \emph{extendable} --- to accommodate the newly-created libraries, 
    \emph{updatable} --- to include the latest library versions,
    and \emph{traceable} --- to track all previous revision details.

\end{itemize}    

\noindent
Unfortunately, 
    the existing TPL databases~\cite{Centris, TPLite, OSSFP} fall short of the necessary standards as they are neither comprehensive nor specific, and they lack efficient mechanisms for updates to include additional whitelists. 
When updating these databases, 
    it is imperative to reconstruct them by meticulously repeating all of the steps, including the library
    collection and redundancy elimination.
This process is time-consuming,
    particularly because  redundancy elimination typically incurs higher costs, 
    requiring approximately 100 hours \cite{Centris,OSSFP}.
Therefore, 
    a powerful and adaptive database is expected to be built.

\vspace{0.05cm}
\noindent
\textbf{Detection of TPL reuse.}
\label{tpl_reuse_overview}
Apart from the platform diversity,
    each TPL includes a variety of functions to serve different purposes, 
    allowing developers to choose specific APIs that align with their unique requirements. 
For instance, 
    \tool{coreMQTT}~\cite{coremqtt} provides two APIs, \func{MQTT\_ProcessLoop()} and \func{MQTT\_ReceiveLoop()},
    for receiving packets from the transport interface iteratively. 
If it is not necessary to keep the receiving portal alive, 
    \func{MQTT\_ProcessLoop()} can be selected; otherwise, \func{MQTT\_ReceiveLoop()} should be chosen.
Therefore, 
    it is necessary to check the functions invoked in the target program by matching them with the functions declared in the libraries. 
Nonetheless, 
    such similarity-based determination is heuristic 
    and highly relies on a threshold established through manual observation.
A high threshold might miss some cases;
    whereas a low threshold leads to many false alarms.
Due to the diverse infrastructure, 
    developers may modify some functions in TPLs to fit their code structure better, 
    rather than reusing TPLs exactly, 
    which complicates the similarity comparison process as well. 
Hence, 
    additional details such as birth times of when each version of TPLs was created are employed as auxiliary information~\cite{Centris}. 
Relying on single auxiliary information may be insufficient to validate the complicated dependencies among TPLs and the dependent software. 

\begin{figure}[t] 
    \centering 
    \includegraphics[width=\linewidth]{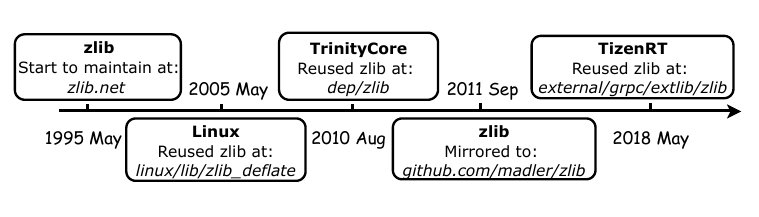} 
    \caption{Example of zlib Maintenance and Reuse} 
    \label{fig:motivation_example1} 
\end{figure}

Considering the real-world TPL reuses demonstrated in Figure~\ref{fig:motivation_example1} as an example, 
    it illustrates the dependencies among \tool{zlib}, \tool{Linux}, \tool{TrinityCore}, and \tool{TizenRT}, and the birth time of each project. 
Specifically, 
    \tool{zlib} was created at \emph{zlib.net} in May 1995 and was mirrored to GitHub in September 2011.
\tool{Linux}, \tool{TrinityCore}, and \tool{TizenRT} separately reuse \tool{zlib} for compression and decompression; thus three projects share some common functions that are invoked from \tool{zlib}.
Since \tool{zlib} was initially released on its own websites and only appears on GitHub later than both \tool{Linux} and \tool{TrinityCore}, 
    the validation through function similarity comparison and birth time ranking ~\cite{Centris} consequently wrongly indicates that \tool{TizenRT} reuses functions from both \tool{Linux} and \tool{TrinityCore}.

\begin{mdframed}[backgroundcolor=gray!20, linewidth=0pt, innerleftmargin=10pt, innerrightmargin=10pt, innertopmargin=10pt, innerbottommargin=10pt]
\textbf{Conclusion 1:} The comparison process should be optimized by incorporating sufficient and accurate TPL information for effective TPL reuse detection.
\end{mdframed}

\vspace{0.05cm}
\noindent
\textbf{1-day vulnerability detection within TPLs.}
\label{1-day_overview}
The integration of TPLs within the development cycle of OSS not only offers convenience for implementing common features but also introduces security and privacy hazards by importing security issues.
Some of these vulnerabilities arise from integrating the vulnerable versions of TPLs~\cite{vuddy, Firmsec, LibScan, v1scan, Holmes}, 
    while others occur because of the violations of specific usage requirements~\cite{LibScout, OSSPolice, zhao2023uvscan}.
1-day vulnerabilities are typically caused by the former.

Similar to TPL reuse detection,  
    some vulnerable features are required for syntactical and semantic comparison to recognize the potential vulnerabilities introduced by the reused TPLs. 
According to the types of reused TPLs, 
    the exact TPL reuses can be commonly analyzed via feature comparison~\cite{OSSPolice,Firmsec, LibScan} or even solely through the comparison of TPL versions.

Alternatively, 
    TPL reuse can be customized by modifying specific segments to achieve functional objectives. 
While some studies have considered customized reuses, 
    they heavily rely on semantic analysis at a broad level to pinpoint the common features that appear in both vulnerable code and patches (e.g., variable names~\cite{vuddy}, core lines of statements~\cite{v1scan},coarse-grained function abstraction~\cite{mvp}). 
Many false alarms are consequently reported.
Hence, 
    it is essential to differentiate the critical statements and operations that are the root causes of a vulnerability for vulnerability identification.

\begin{figure*}[!t]
      \centering
      \includegraphics[width=\textwidth]{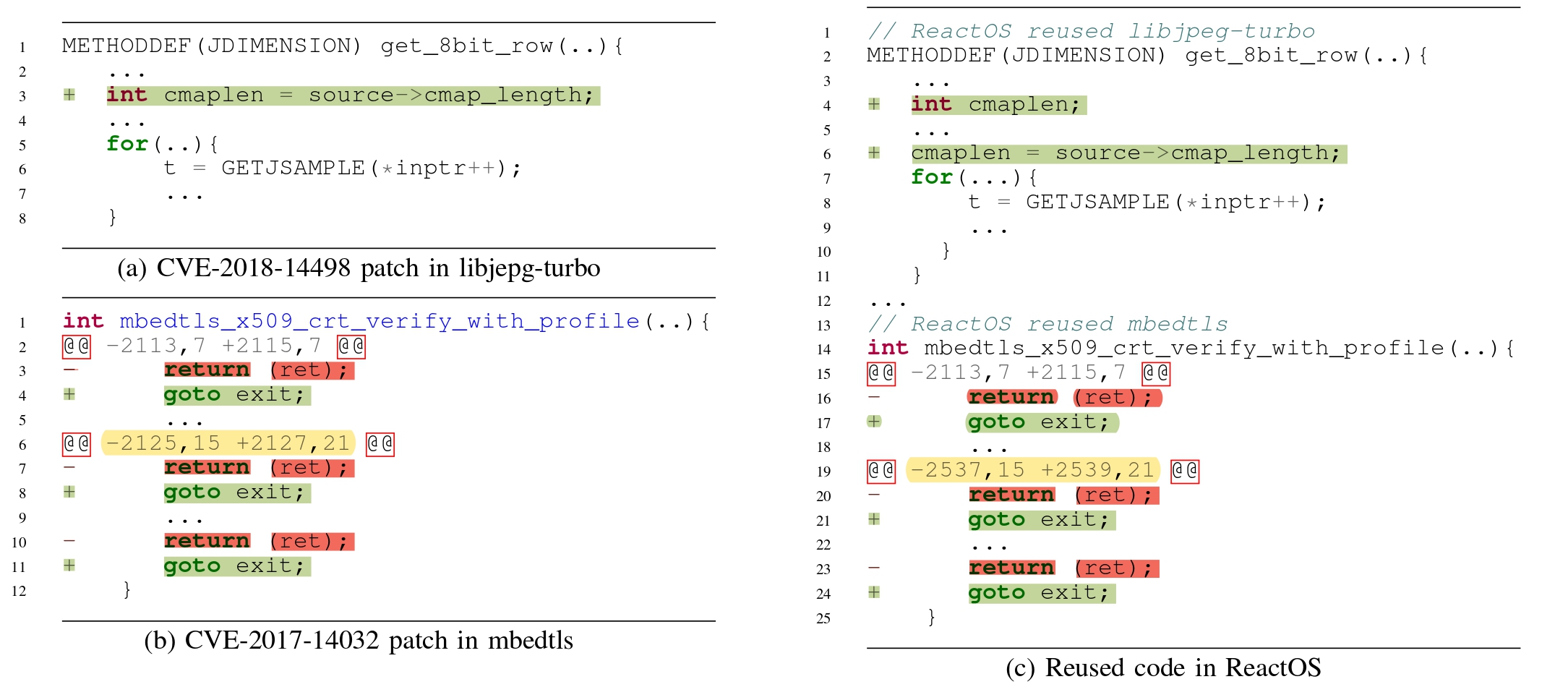}
      \caption{ReactOS patching of CVE-2018-14498 and CVE-2017-14032. Due to custom reuse, the ReactOS patch differs from the official one in statement format and the line numbers where the patches are applied.}
      \label{fig:motivation_example2}
\end{figure*}

Considering the TPL reuses in \tool{ReactOS} (shown in Figure~\ref{fig:motivation_example2}) as an example, 
    \tool{ReactOS} originally reused vulnerable TPL versions of \tool{libjepg-turbo} and \tool{mbedtls}. 
Then it patches the vulnerabilities in a custom way. 
While the statement contents remain the same, 
    the line numbers where the statements are located and the statements format changed, 
    such alteration challenges the existing approaches~\cite{vuddy}~\cite{v1scan} to accurately identify the custom patches.

\begin{mdframed}[backgroundcolor=gray!20, linewidth=0pt, innerleftmargin=10pt, innerrightmargin=10pt, innertopmargin=10pt, innerbottommargin=10pt]
\textbf{Conclusion 2:}
Selecting critical vulnerable features that contribute to the generation and mitigation of a vulnerability can enhance the effectiveness and efficiency of 1-day vulnerability detection.

\end{mdframed}

\subsection{CVE and Commit Analysis}
Software assessment relies on publicly available information on vulnerabilities and security patches,
    which is typically provided by CVE~\cite{CVE} and NVD~\cite{NVD}.
However, 
    many vulnerabilities disclosed by CVE/NVD come with either vague security patch information or no patch information whatsoever~\cite{vcmatch}. 

Commit ranking is a common process~\cite{patchscout, vcmatch, uncovering, vfcfinder} of labeling direct references between security patch-related commits and vulnerabilities.
To identify the patch commit of a vulnerability, 
    specific sets of features related to commit candidates and the vulnerabilities (e.g., vulnerability location, vulnerability identifier) are collected.
Then, a pre-trained ranking model is used to estimate the relevance between each commit and the vulnerability. 
The most relevant commits related to security patches are supposed to be prioritized. 
Nonetheless,
    the ranking results highly rely on the quality of the selected correlation features and the information collected from CVE/NVD, 
    which are typically manual processes.
Although the manual effort is a one-time cost, 
    the variable quality of the data collected by humans can affect the accuracy and reliability of the commit ranking. 
Driven by the fact that public LLMs such as \tool{GPT-4.0} have undergone extensive training on a
vast array of data accessible online, including documentation,
blogs, and forums, 
    LLMs have become highly proficient in grasping the semantic meaning of natural language~\cite{gpt4report}.

\begin{mdframed}[backgroundcolor=gray!20, linewidth=0pt, innerleftmargin=10pt, innerrightmargin=10pt, innertopmargin=10pt, innerbottommargin=10pt]
\textbf{Conclusion 3:}
LLM can assist in natural language processing to analyze commits and CVE descriptions semantically and syntactically.
It minimizes manual efforts, 
    enhancing the data quality in identifying security-related information and assessing the implications of code changes.  

\end{mdframed}

\section{Overview}
\label{sec:overview}

\begin{figure*}[t]
  \centering
  \includegraphics[width=\textwidth]{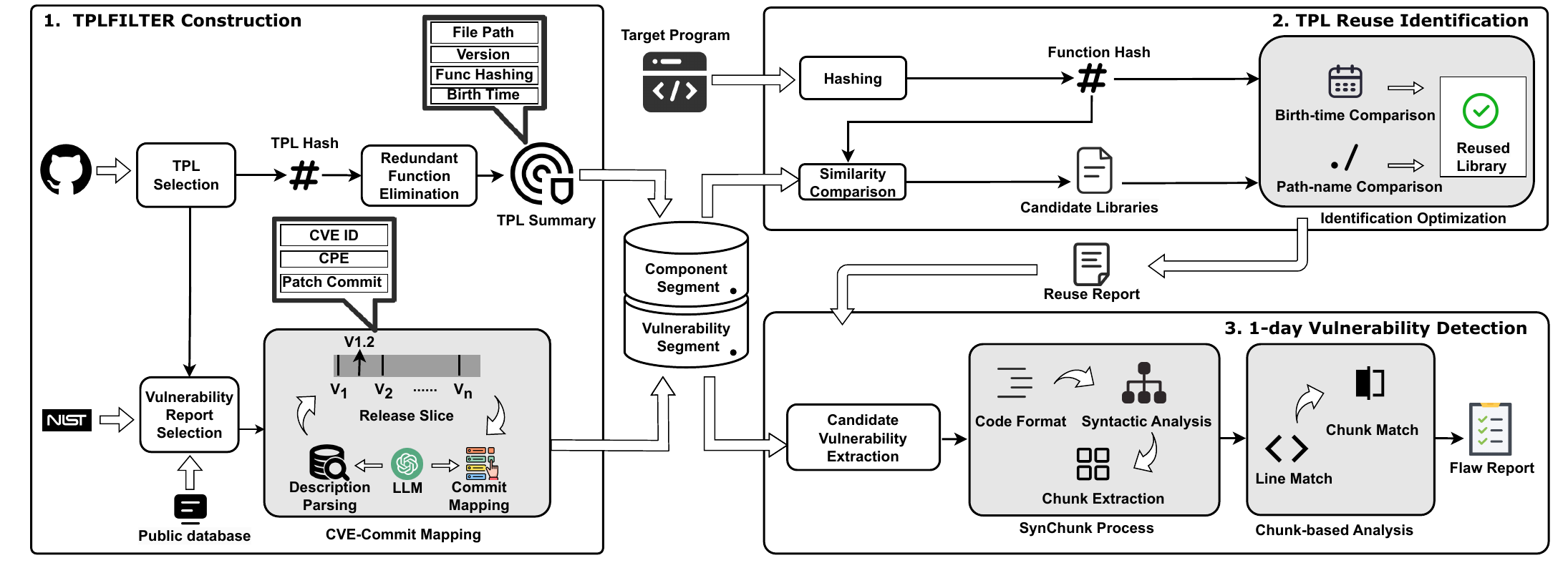}
  \caption{Workflow of \sysname}
  \label{fig:workflow}
\end{figure*}

Figure~\ref{fig:workflow} illustrates the workflow of \sysname for detecting 1-day vulnerabilities introduced by reusing TPLs. 
\sysname consists of three phases, \emph{\tplname Construction}, \emph{TPL Reuse Identification}, and \emph{1-day Vulnerability Detection}.

\noindent
\textbf{\tplname Construction.}
\sysname employs \tplname to construct a unique database tailored specifically for the target platform. 
It is comprised of two segments, the component segment, and the vulnerability segment. 
The component segment contains the TPL details (e.g., TPL names, TPL versions, and code information) while the vulnerability segment includes information on vulnerabilities that have previously or currently existed in the previous and current versions of each TPL, respectively. 

\noindent
\textbf{TPL Reuse Identification.}
Given a target program,
    \sysname conducts a function-level TPL reuse detection. 
By extracting the function hashing of all functions in the target program,
    \sysname identifies the TPLs and their versions that have been reused by the target program via function-based similarity comparison. 
To eliminate the false alarms driven by the custom TPL reuses,
    \sysname analyzes the dependencies among TPLs to optimize the results and creates a TPL reuse report. 

\noindent
\textbf{1-day Vulnerability Detection.}
As TPLs are reused in two different ways, i.e., exact reuse and custom reuse, 
    \sysname analyzes each type of reuse by utilizing version-based matching and chunk-based analysis, respectively. 
Through version-based matching, 
    \sysname specifically identifies the functions that are reused exactly and verifies whether these functions are vulnerable by searching for the official patches provided by the TPL contributors. 
In cases of custom reuses, 
    \sysname re-generates the modified functions into chunk representatives and analyzes the code modifications within each chunk. 
It can determine whether the modifications introduce any vulnerabilities at a granular level.  
As \tplname includes the official patches for certain vulnerable TPLs, 
    \sysname further provides fix suggestions based on the reported vulnerabilities. 

\section{\sysname}
\label{sec:detection}


\subsection{\tplname}
\label{approach:multi_db}
To detect 1-day vulnerabilities introduced by TPL reuses, 
    \sysname must 1) identify the reused TPLs in the target program;
    2) verify whether the reused TPL version is vulnerable; 
    3) suggest if any potential patch is available.
Therefore, 
    \sysname is required to construct a database, 
    containing the TPLs commonly used by a platform and the corresponding vulnerable and patch information as a detection reference.

\subsubsection{TPL Selection} 
As TPLs vary across different platforms, 
    \sysname executes \tplname to construct a database containing TPLs specific to the targeted platform (e.g., IoT firmware, Android, iOS).
Since most open-source projects are maintained and operated on \tool{GitHub}, 
    \tplname first collects all prevalent libraries from GitHub or existing TPL databases, such as \tool{OpenWRT}~\cite{openwrt} and \tool{Awesome Android}~\cite{awesome-android-libraries}.
To select TPLs specific to the target platform, 
    we manually created a keyword list of the most commonly used keywords for describing the target platforms. Through the keyword list, \tplname performs keyword matching to determine if the titles, tags, and descriptions of each collected TPL contain any of the specified keywords. 
Additionally, 
    \tplname scans the project metadata, official websites of projects, and \tool{README} files to identify the keywords (e.g., system, server, firmware) for non-library project exclusion.

\subsubsection{Component Segment Construction} 
\label{component_construction}
Within the database, 
    \tplname gathers detailed information on each selected TPL to construct the component segment.
GitHub repositories and official websites may provide a variety of information about the TPL, 
    including details about their functions, file names, license information, and the creation time of each library (i.e., birth time). 
However,
    we observed that file names are not unique to every TPL and licenses are not formally organized by all target programs, 
    which may negatively affect the effectiveness of TPL reuse detection. 
Alternatively, 
    \tplname analyzes functions declared in each TPL and their birth times to build the database, 
For efficiency, rather than storing all the complicated details of functions, 
    \tplname calculates a hash value for each function and stores them in the component segment. 
    

For each selected TPL, 
    \tplname clones them from GitHub and lists all the published versions through \func{git tag}. 

To minimize false alarms arising from string-matching in the handling of custom patches,
    \tplname employs LSH to process each function. 
LSH is a fuzzy hashing technique that hashes similar input items into the same ``buckets" with high probability, 
    thereby enhancing the accuracy of data comparison~\cite{jafari2021survey}.
Hence, 
    \tplname analyzes each TPL version and extracts all functions with \tool{ctags}. 
Then, it calculates the hash value of each function.

Simultaneously, 
    it executes \func{git log} to identify the birth time of each function from the commit histories.
Each function is then represented as $fc = <H, Birth>$, 
    where $H$ is the hash value of the function and $Birth$ refers to its birth time.
Hence, 
    each version of a TPL is comprised of the functions, represented as $FC = \{fc(i)|1 \leq i \leq n\}$, where $n$ is the maximum number of functions included in the TPL version.
As a TPL $A$ may invoke functions from another TPL $B$ (i.e., $A$ depends on $B$),
    such dependencies may lead to redundant comparisons when analyzing function similarities. 

Although some previous works (e.g., \tool{Centris}~\cite{Centris} and \tool{OSSFP}~\cite{OSSFP}) also utilize LSH to calculate hash values,
    these works identify redundant functions hash value similarity comparison rather than hash value exact match identification. 
Such schemes become inefficient when comparing functions on a large scale.

To eliminate redundant comparisons, 
    \tplname employs a hashing-index based elimination method that narrows down the scope by searching for functions with identical hash values. 
Specifically, 
    \tplname first identifies the functions $fc(i)$ and $fc(j)$ by comparing hash values $H(i)$ and $H(j)$.
Among all the functions with the same hash values, 
    it selects the function with the earliest birth time and removes the other functions. 



\subsubsection{Vulnerability Segment Construction}
To support the determination of whether the reused versions of TPLs contain any vulnerabilities, 
    \tplname collects all vulnerabilities associated with each version of the TPLs in the component segment.
As some developers may patch vulnerabilities on their own without updating TPL versions,
    \tplname assesses security patches, instead of TPL version, to identify whether a vulnerability affects the target program.  
Hence, \tplname collects security patches of each collected vulnerability, and compiles these vulnerabilities and their respective security patches in the vulnerability segment. 
    

\vspace{0.05cm}
\noindent
\textbf{Vulnerability detail collection.} 
As the majority of vulnerability reports are typically listed on \tool{CVE} and \tool{NVD},
    \sysname crawls the vulnerability reports from these websites. 
Within these reports, each vulnerability is assigned a unique CVE ID for identification and is described using a CPE (Common Platform Enumeration) to specify the software versions affected.

Given each TPL stored in the component segment,
    \sysname first filters the vulnerability reports whose descriptions contain the TPL name at a coarse level by invoking the NVD API with \func{keywordSearch}\footnote{https://services.nvd.nist.gov/rest/json/cves/2.0?keywordSearch=\{*\}, 
    where element \func{*} refers to the name of the target TPL.} .
Nonetheless, 
    vulnerability descriptions encompass not only the names of vulnerable libraries but also the names of other software impacted by these vulnerable libraries, rendering keyword searches less precise.
Therefore, 
    \sysname conducts a CPE-guided matching to filter vulnerability reports for effective vulnerability identification. 
Specifically, to determine whether a vulnerability affects a target TPL,
    it retrieves the CPE information in the vulnerability report and checks whether the target TPL name is the substring of the CPE string.
If the target TPL name is matched, 
    \sysname regards that the vulnerability affects the target TPL and records the corresponding information of the CVE ID, its CPE, and the vulnerable TPL version(s). 
Typically, 
    CPE manages the vulnerable software versions in two ways: through enumerations or version intervals.
For enumeration, 
    \sysname directly extracts all vulnerable version numbers.
In the case of interval specification, 
    \sysname first obtains and ranks all versions of a TPL in ascending order, and then determines the start and end versions that fall within the specific vulnerable version interval. 

Unfortunately, 
    we inspected that some vulnerability reports are informally written, 
    thus these vulnerabilities might be overlooked by keyword searches and CPE matching.
To address such issues,
    \sysname extracts additional vulnerability reports from public databases including \tool{SNYK}~\cite{Snyk_Vulnerability_Database} and the official websites of TPLs~\cite{ffmpeg}~\cite{curl} for further detail matching.

\vspace{0.05cm}
\noindent
\textbf{Security patch collection.} 
\sysname collects security patches for each vulnerability to support accurate 1-day vulnerability detection and examine whether a reused TPL in the target programs has been patched.
Although \tool{CVE}/\tool{NVD} includes patch information for some vulnerabilities, 
    a significant portion of the patch information is not updated timely or maintained properly~\cite{cveWithNoPatch, WhereFixes}. 
Alternatively,
    TPLs maintained on \tool{GitHub} frequently release security patches as GitHub commits, 
    which typically include descriptions of vulnerabilities
    as well as detailed code changes.
    
To effectively and comprehensively obtain the vulnerability patches, 
    \sysname constructs an LLM-based multi-sliced patch searching approach to locate the specific security patches of each vulnerability.  
Given the CVE description maintained by \tool{CVE}/\tool{NVD} and the repository commits maintained by GitHub of each TPL, 

\sysname proceeds with the following four steps:


\noindent
\textbf{1) LLM-based description parsing:} 
\label{subsec:llm_des_parsing}
The vulnerability detail provided by \tool{CVE}/\tool{NVD} typically consists of the vulnerability description (e.g., vulnerability type, vulnerable files, vulnerable functions, vulnerable variables, and some specific vulnerable features), 
    and CPE describing the affected TPL information 
    (e.g., vendor(s) of the TPL,  TPL name and versions). 
Correspondingly,
    \sysname employs a LLM by invoking \tool{GPT3.5} to parse the vulnerability description and then extracts the vulnerable elements affected by the vulnerability, i.e., vulnerable files, vulnerable functions, and vulnerable variables.



\noindent
\textbf{2) Slice-based commit filtering:}
\label{subsec:slice_based}
As a vulnerable TPL repository may contain numerous commits that cover document changes,
    routine common bug fixes, feature modifications, and vulnerability patches, 
    it is time-consuming and error-prone to analyze those commits one by one. 
Therefore, 
    \sysname conducts a date-specific commit slicing to exclude the commits confirmed to be irrelevant to the vulnerabilities at a coarse-grained level.
Specifically, 
    \sysname first identifies all the vulnerable versions from the vulnerability description and CPE information. 
Since the version released immediately after the last vulnerable version is typically the first patched version of a vulnerability,
    \sysname extracts the timestamps of when the last vulnerable version and the first patched version were published.
Correspondingly, 
    all the commits generated within the time range between these two timestamps are considered as potential patch commits.
Then, 
    it divides the potential patch commits into multiple slices based on average partitioning (i.e., each slice contains $k$ commits.
Within each slice $i$, 
    \sysname calculates a code diff \textit{diff$_i$} between the first and the last commits.
If \textit{diff$_i$} contains any of the vulnerable elements, 
    \sysname labels the slice $i$ as a candidate slice, 
    indicating that at least one commit within this slice contains modifications related to vulnerability-specific code elements.
Otherwise, 
    it continues analyzing the next slice $i+1$.

\noindent
\textbf{3) Candidate commit selection:}
\label{subsec:candidate_commit}
After locating the candidate slice, 
    \sysname performs fine-grained analysis to select the candidate commits that are related to the vulnerability. 
It compares vulnerable code with each patch to generate a code diff.
If the code diff includes any of the vulnerable elements,
    \sysname labels the commit as a candidate commit and advances to step (4) for further confirmation.




\noindent
\textbf{4) LLM-based patch commit mapping:} 
\label{llmmss:step3} 
\sysname employs LLM to confirm the patch commits.
On one hand, some commits within the candidate commits may also modify the vulnerable elements yet are not the patch regarding the target vulnerability, requiring \sysname to exclude these commits;
on the other hand, 
    some CPE information provided by NVD is incorrect~\cite{idaffectedversion} which may lead to false alarms.
To address the challenges,
    \sysname takes as input the CVE description and each candidate commit (i.e., commit description and the modified code), and invokes \tool{GPT4.0} to infer whether the CVE and the commit are relevant.
It is important to note that only minor candidate commits are needed to be analyzed here, 
    thus \tool{GPT4.0} incurs minimal costs to calculate the correlations.

A detailed example of patch commit mapping is illustrated in Appendix~\ref{appendix_b}.

\subsection{TPL Reuse Identification}
To determine what TPLs are reused in each target program, 
    \sysname takes as input the component segment of the database to check whether these TPLs are invoked.
It first employs an LSH algorithm to generate a list of TPL candidates that the target program may use. 
Subsequently, 
    it refines the identification process by incorporating additional information for further confirmation. 



\subsubsection{Candidate Library Detection}
\label{subsec:candidate_detection}
As TPLs may be reused either exactly or in a custom manner, 
    relying solely on matching function names and code statements may overlook such custom cases.
Hence,
    \sysname leverages the LSH algorithm to conduct code similarity analysis.

It extracts all files along with their paths from the target program and then proceeds with these files one by one. 
Then,
    \sysname divides each file within the target program into target function snippets based on the declared functions. 
Given each function snippet, 
    it uses Python \tool{TLSH}, 
    an LSH-based fuzzy matching library, 
    to generate a unique hash value, 
Each target function snippet $i$ can be represented by a 2-tuple: $ft(i) = <Hash(i), Func\_path(i)>$,
    where $Hash$ is the hash value of the target function snippet $i$ and $Func\_path$ represents the file path where the function snippet $i$ is defined.

\sysname then computes the similarity between the target function snippets and TPL functions in the component segment.
For each target function snippet $i$, 
    \sysname compares it with each function $j$ from the TPL iteratively.
In particular,
    it takes as input $Hash(i)$ and $H(j)$ and leverages \tool{TLSH} to calculate a similarity score between functions $i$ and $j$.
If the similarity score falls below a hash threshold $TH_{hash}$, 
    the function pair of $i$ and $j$ is considered similar.
After all target function snippets are analyzed, 
    \sysname calculates the total number of similar function pairs. 
When the total number exceeds a similarity threshold $TH_{sim}$, 
    it represents that the corresponding TPL may be used.
Since different versions of a TPL may vary significantly, 
    \sysname identifies the version with the highest number of similar function pairs as a TPL prevalent version and then captures the information of the TPL (i.e., TPL name, hash value of the function in the TPL, TPL version, and the file path where the TPL is reused in the target program) to store in the candidate TPL list.
It is important to note that when multiple TPL versions have the same number of similar function pairs, 
    \sysname randomly selects one version as the TPL prevalent version.
\sysname iteratively analyzes all functions in the target program and collects a list of candidate TPLs that are potentially reused.

\subsubsection{Identification Optimization}
\label{idop}

The common similarity-based approaches, such as \tool{Centris}, may produce a significant number of false positives because 1) TPLs with similar functionalities may employ the same functions; 
    2) the inter-dependencies among TPLs may result in a high degree of function overlap between TPLs. 
To accurately select which TPL is reused, 
    \sysname leverages auxiliary information, i.e., file paths and birth time, to optimize the candidate list.

Some existing approaches (e.g., \tool{SNYK}~\cite{Snyk_Vulnerability_Database}) use TPL licenses to identify reuses because the target program must declare a TPL license either in a LICENSE file or through an annotation when reusing.
However, 
    we observed that these TPL licenses are usually written inconsistently and informally, 
    without adhering to standards. 
\sysname may overlook the declared TPLs when using such information. 
Hence, 
    \sysname examines file paths illustrating where TPLs are reused in the target program and the birth time of each TPL as the indicators.
Generally,
    the target program places code that reuses TPLs under a separate directory 
    and then names the directory with a title similar to the TPL name. 
The birth time is used to pinpoint the first invocation of the reused function,
    thereby eliminating the redundant TPL dependencies involved during function comparison. 




To confirm the reused TPLs, 
    \sysname first groups TPL candidates together when they share the same file path in the target program. 
For the TPL candidates in the same group,   
    a path tokenizer is used to divide each file path into tokens based on the separators (e.g., space, backslash, and colon).
Then \sysname compares each token with the name of each TPL candidate and calculates a Jaccard similarity score~\cite{niwattanakul2013using}, and selects the TPL candidate(s) with the highest Jaccard similarity score as the confirmed reused TPL.
Note that a file may invoke functions from multiple TPLs, 
    thus several TPLs can have the same Jaccard similarity score.

\sysname further addresses the inheritance cases where several TPLs share the same score to locate the confirmed reused TPL.
First, 
    \sysname gathers the remaining TPL candidates sharing the same hash value $H$ into a group, 
    indicating that these TPLs are related. 
Within each group, 
    \sysname compares the birth time of each TPL and identifies the one with the earliest birth time as the parent TPL of the group (i.e., the confirmed TPL of the target program). 
All the remaining TPL candidates in the group are then removed. 

For each target program, 
    \sysname eventually generates a TPL reuse report containing all the confirmed reused TPLs.

\subsection{1-day Vulnerability Detection}
\label{approach:flaw_detection}

When the TPL reuse report discloses that a vulnerable version of TPL has been reused, 
    \sysname then determines whether any vulnerable functions are being invoked exactly or in a custom way. 
Hence, 
     \sysname implements a dual analysis strategy at the function level, 
     utilizing version-based and chunk-based analyses to distinguish exact and custom reuses, respectively.

\subsubsection{Version-based analysis}
By inspecting the reused TPLs, 
    we found that some developers may reuse a vulnerable TPL version, but patch the vulnerable code snippets themselves.
Therefore,
    \sysname utilizes a diff-inspired version-based analysis to examine a TPL reuse from three aspects: 1) whether reuse is exact or customized in terms of TPL reuses; 2) whether the reused TPL version is vulnerable; and 3) whether the exploited vulnerabilities are patched.

In detail,
    \sysname first takes the TPL reuse report as input to extract the reused TPL and its version.
According to the TPL name and the version, 
    it queries the vulnerability segment of the database to verify if the reused version of the TPL was linked to any CVE report. 
If a CVE report is identified, 
    \sysname extracts the vulnerable and patched version of the code related to the CVE, 
    and then utilizes \func{diff} to locate the code diff \textit{Diff$_{vp}$}. 
Subsequently, \sysname analyzes the \textit{Diff$_{vp}$} to determine if the modified items in \textit{Diff$_{vp}$} are functions or global declarations (e.g., global variables, structures, and macros). 
For functions, 
    \sysname extracts and records the entire function of both the vulnerable and patched code as the vulnerable and patched functions, respectively; 
    while for global declarations, \sysname records them as the vulnerable and patched global declarations.
Then, within the target program, leveraging \tool{ctags},
    \sysname specifically labels each code snippet as either a function or a global declaration and compares them with vulnerable/patched functions/declarations, respectively.





As a result, 
    \sysname distinguishes TPL reuses in the target program into four groups and processes each group with function-level granularity, ensuring that unused functions in the TPLs do not affect the detection accuracy:


\noindent
\textbf{G1: No vulnerable reuse.} It indicates a target program reuses the vulnerable version of a TPL \emph{without involving any} vulnerable functions or vulnerable global declaration.
\begin{itemize}
\item \textbf{Analysis:} When all the target code snippets are exactly equal to the patched function or the patched global declaration, 
    \sysname regards the target program as G1. 
The target programs in this group are considered secure.
\end{itemize}
    
\noindent
\textbf{G2: Vulnerable global declaration reuse.} It represents that a target program only reuses vulnerable global declarations.

\begin{itemize}
\item \textbf{Analysis:} When any of the global declarations in a target code snippet exactly equals to that of the vulnerable global declarations, 
    \sysname regards the target program as G2. 
\item \textbf{Confirmation:} \sysname employs line matching to 
identify the vulnerable declarations. 
If a global declaration that is deleted from the vulnerable TPL continues to appear in the target code snippet, 
    \sysname considers that a vulnerability is identified.
Furthermore,
    when a global declaration added to the patched TPL does not appear in the target code snippet, 
    \sysname considers the target program as vulnerable. A 1-day vulnerability is then identified.
    

\end{itemize}
   
\noindent
\textbf{G3: Exact vulnerable reuse}: It represents that a target program exactly reuses the vulnerable function of a TPL without providing any patches.
\begin{itemize}
\item \textbf{Analysis:} \sysname labels the target program as G3 if any target functions are equal to the vulnerable functions.

\item \textbf{Confirmation:}
\sysname utilizes \tool{TLSH} to compute the hash values of the vulnerable function and the target function snippet. If any function in the target program shares the same hash value as the vulnerable function, \sysname considers it unpatched and classified to G3. A 1-day vulnerability is then identified.

\end{itemize}

\noindent
\textbf{G4: Custom reuse:} It represents that a target program reuses the vulnerable or patched functions with custom modifications. 

\begin{itemize}
\item \textbf{Analysis:}  \sysname labels the target program as G4 if any target functions are similar to the vulnerable or patched functions.
If the similarity score for any pair of functions exceeds $TH_{sim}$, \sysname classifies the target program as G4. 
\item \textbf{Confirmation:} \sysname uses \tool{TLSH} to compare the hash values of each function in the target program with those of the vulnerable function. 
Since it involves custom reuse, it is uncertain whether the target program is secure; 
    thus, \sysname confirms the vulnerable custom reuses via a chunk-based analysis.

\end{itemize}

\subsubsection{Chunk-based analysis}

To preserve the semantic and syntactic information that are overlooked by function~\cite{vuddy} and line matching~\cite{v1scan},static analysis has traditionally been used to address this issue, but it comes with significant challenges. First, static analysis requires compilable code, which is problematic because patches are often provided as isolated snippets that cannot be compiled directly. Second, static analysis demands an extensive environment setup for compilation, a process that is both time-consuming and resource-intensive.

   \sysname utilizes chunk-based code analysis to examine custom reused functions, striking a balance between efficiency and accuracy in semantic information extraction without the reliance on compilation. For these customized functions, \sysname verifies whether these custom modifications provide the same security functionality as those in the official patches released by TPL contributors. Specifically,in chunk-based analysis, a \emph{chunk} represents a group of modified lines focused on a specific functionality. In a given function, the lines within the same chunk are either governed by the same control structures (e.g., \func{if-else}, \func{while}, \func{for}) or operate on the same set of variables. This design preserves both the semantic meaning and contextual relationships embedded within the patch code, ensuring accurate assessment of custom adaptations.



Taking as input the information of CVE ID, vulnerable code snippets, patched code snippets, \textit{Diff$_{vp}$}, and target code snippets, 
    \sysname proceeds with the following steps:

\vspace{0.05cm}
\noindent
\textbf{Chunk construction.}
When reusing TPLs, the target program might alter the format of the reused code. Thus, \sysname first standardizes all code snippets, including target code snippets, vulnerable code, and the patched code.
According to the LLVM coding standard\footnote{LLVM Coding Standard: https://llvm.org/docs/CodingStandards.html},
    \sysname removes unnecessary entities such as leading spaces, comments, and constant strings.

After normalization, 
    \sysname separately initiates code diff among the target, vulnerable, and patched code snippets.
The comparison correspondingly yields three code diff (i.e., \textit{Diff$_{vt}$}, \textit{Diff$_{pt}$}, and \textit{Diff$_{vp}$}), where \textit{Diff$_{vt}$} is the code diff between the target and vulnerable code snippets, \textit{Diff$_{pt}$} is the code diff between the target and patched code snippets, and \textit{Diff$_{vp}$} is the code diff between the vulnerable and patched code snippets.
Each line in the above code diff is treated as an individual chunk.
We manually predefined 40 regular expressions, 
    covering all types of statements (e.g., call expressions, binary operations). 
Using these regular expressions,
    \sysname then extracts the variables and operations involved in each chunk.
It further adopts a disjoint-set union algorithm to identify the data-dependent chunks when 1) the chunks are part of the same control block without any nested control structures, or 2) chunks share the same variable.
Eventually, 
    all related chunks are merged,
    while those unmerged chunks remain independent from each other.

\vspace{0.05cm}
\noindent
\textbf{Vulnerability detection.} 
After constructing the chunks, 
    \sysname uses chunk matching to determine if variables in the target code snippet have been patched. 
Initially, 
    \sysname conducts disjoint checking by pairwisely comparing each chunk in \textit{Diff$_{vt}$} and \textit{Diff$_{pt}$} with the chunks from \textit{Diff$_{vp}$} to determine whether the patches have been applied to the target program. 
If both sets of \textit{Diff$_{vt}$} and \textit{Diff$_{pt}$} share the same variables with \textit{Diff$_{vp}$}, 
    \sysname further performs line matching.
A vulnerability is considered ``patched'' if all lines in \textit{Diff$_{vp}$} appear in \textit{Diff$_{vt}$}, 
    but disappear in \textit{Diff$_{pt}$}. 
Otherwise, 
    \sysname further executes operation match.
When all operations in \textit{Diff$_{vp}$} appear in \textit{Diff$_{vt}$}, 
    but disappear in \textit{Diff$_{pt}$}, 
    it indicates that the vulnerability has been patched via customized modifications. 
If both matching conditions are failed,
    a 1-day vulnerability is considered to be ``found''.

\begin{figure}[b] 
  \centering
  \includegraphics[width=1\linewidth]{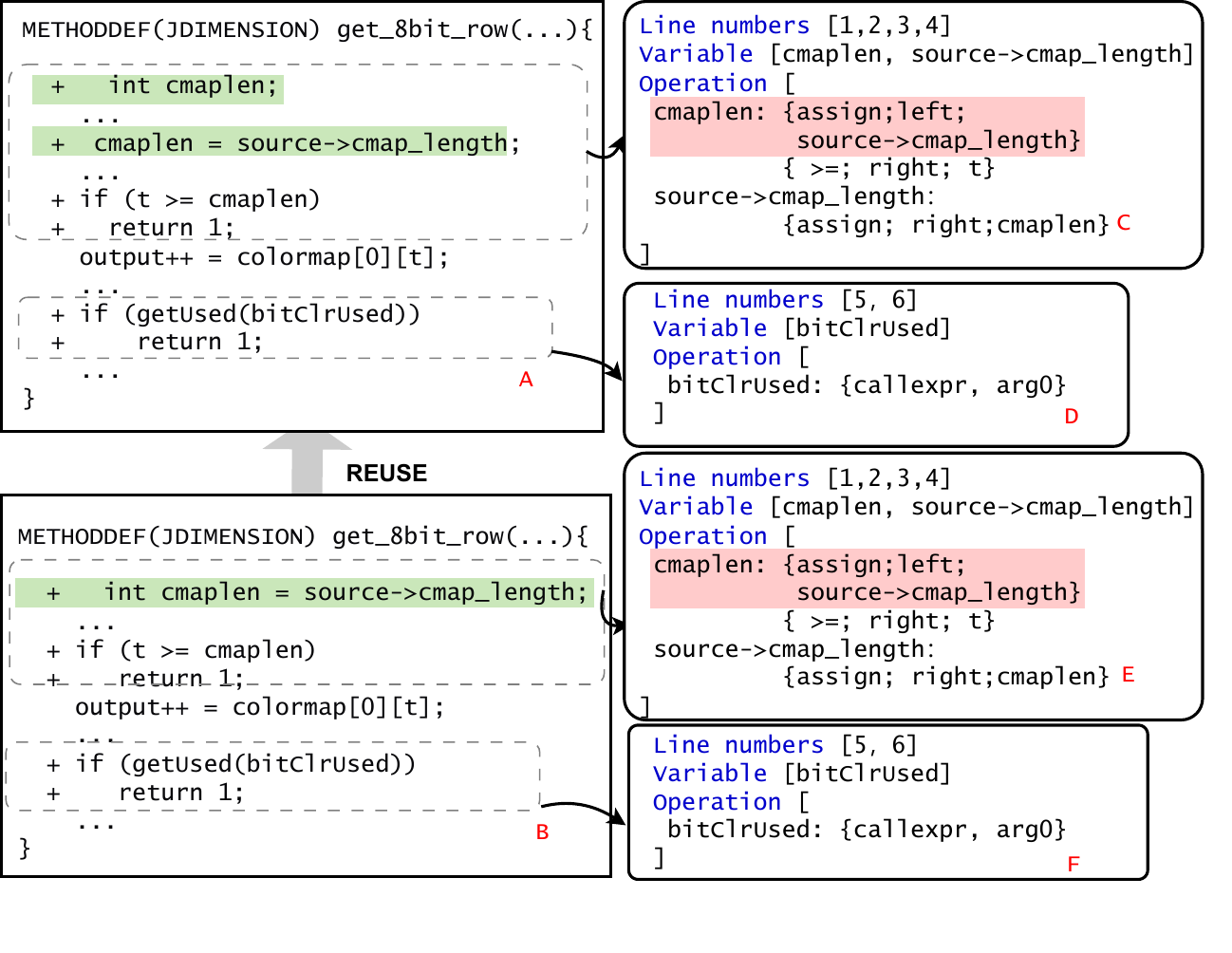}
  \caption{Chunk extraction and match}
  \label{fig:chunk}
\end{figure}

A detailed example for chunk-based detection is depicted in Figure \ref{fig:chunk}. The code snippet A at the top represents the patch code, while code snippet B at the bottom shows the customized patch.

To construct chunks, \sysname initially treats each added line in the patch and reused code as an individual chunk. Next, \sysname merges these chunks based on semantic relationships and context. For instance, in the patch code snippet A, the first four lines are merged into a single chunk because they share the same variable, \emph{cmaplen}. Similarly, the last two lines are merged into another chunk as they are governed by the same control block.

After constructing the chunks, \sysname start to process chunk matching. In the above example,
    Chunk C pairs with chunk E, 
    and chunk D pairs with chunk F.
Code B is classified as patched, 
    since though the line match was unsuccessful 
    (indicated by the highlighted lines on the left side),
    the operation match succeeded 
    (indicated by the highlighted lines one the right side).
This scenario of custom patch may lead to false positives in other systems like \tool{V1SCAN}, 
    but \sysname can accurately identify the patch.


After processing all potential vulnerabilities for the target program through version-based and chunk-based analysis, \sysname will generate a comprehensive vulnerability report. This report not only includes the CVE-ID of the vulnerability but also provides the patch commit URL from our vulnerability segment of the database, along with details of the chunks involved in the detection process, to facilitate further patching on 1-day vulnerabilities by developers.

\section{Experiment}
\label{sec:experiment}
We evaluate the performance of \sysname across two aspects: 
    the accuracy of TPL reuse detection, and the effectiveness in detecting 1-day vulnerabilities introduced by the reused TPLs.


\subsection{Experiment Setup}

Although \sysname can be applied to any arbitrary platform for 1-day vulnerability detection,
    we applied \sysname to IoT projects as a representative in our following experiments.
In detail, 
    \sysname collected all prevalent repositories from \tool{GitHub}, \tool{OpenWRT}, \tool{stm32duino}, 
    \tool{awesome-cpp}, 
    \tool{awesome-c}, 
    and \tool{mongoose-os-libs}  in C/C++, 
    and then conducted TPL selection to choose the specific TPLs.
After filtering the repositories via keywords, 
    19,057 libraries were left. 
We labeled these libraries as a database DB$_{kwd}$.
After identifying the dependent  libraries and only retaining the parent TPLs, 
    \sysname eventually built a TPL database DB$_{iot}$ containing 1,872 IoT-specific TPLs,
    which include TPL names, all TPL versions, hash values, and birth times of functions within the TPL to store as the component segment within DB$_{iot}$.
According to the collected TPLs, 
    \sysname explored 5,114 CVE reports, of which, 1,717 CVEs have GitHub patches available.
These CVE reports and patches are stored as the vulnerability segment within DB$_{iot}$.

We evaluated the performance of each component in \sysname against corresponding state-of-the-art works.
To evaluate the database quality, (Section~\ref{subsec:database_eval}), 
    we selected \tool{Centris}, \tool{SNYK} and \tool{VFCFinder}~\cite{vfcfinder}.
For vulnerability detection (Section~\ref{subsec:vul_detect}), 
    we chose \tool{V1SCAN}, 
    \tool{TPLite}\cite{TPLite}, 
    and \tool{SNYK}.
Although \tool{MVP} is also a 1-day vulnerability detection tool, 
    was excluded due to its lower reported performance compared to \tool{V1SCAN}. 
Each tool was implemented by using its latest version as of April 2024, 
    following the instructions provided by developers.


We ran \sysname on a machine with Debian GNU/Linux 12, 32GB RAM and 1TB SSD. 
For patch commit collection, we set the search slice size $k$ at 20.
For similarity comparison,
    we set $TH_{hash} = 30$ and $TH_{sim} = 10\%$ after manual testing to achieve optimal detection results.

\subsection{Database Evaluation}
\label{subsec:database_eval}
We evaluated the two segments (i.e., component segment and vulnerability segment) within the database, DB$_{iot}$, respectively.  

\subsubsection{Component Segment Evaluation}
A suitable database for TPL reuse detection must be \emph{comprehensive}, \emph{specific}, and \emph{maintainable} (refer to Section~\ref{sec:background});
    thus we conducted two experiments to evaluate whether DB$_{iot}$ adheres to these characteristics while ensuring the database size remains relatively small:

\begin{itemize}
\item \textbf{Characteristic comparison.} We compared DB$_{iot}$ with the prevalent library database published recently~\cite{Centris} ~\cite{TPLite}.

\item \textbf{Database integration.}  We integrated DB$_{iot}$ with the state-of-the-art TPL reuse detection tool, \tool{Centris}, to verify if its detection performance is enhanced.
\end{itemize}

\begin{table*}[]
\centering
\caption{Maintainability of Each Database}

\begin{tabularx}{0.87\textwidth}{c|c|cc|c|c|cc}
\toprule[1.5pt]
\multirow{2}{*}{\textbf{Database}} & \multirow{2}{*}{\textbf{TPL Number}} & \multicolumn{2}{c|}{\textbf{Time Cost Per TPL Processing}} & \multirow{2}{*}{\textbf{Storage Total(GB)}} & \multirow{2}{*}{\textbf{Frequency of Comparisons}} & \multicolumn{2}{c}{\textbf{Detection Time}} \\ \cmidrule{3-4} \cmidrule{7-8}
                                   &                                      & \textbf{Clone (s)}        & \textbf{Elimination (s)}       &                                                                                        &                               & \textbf{Exact (s)}   & \textbf{Similarity}  \\ \midrule
\textbf{$DB_{iot}$}                     & 1,872                                & 50.5                      & 0.1                            & 3.5                                                                                    & 9,207.1                       & 2.2                  & 256.4                \\
\textbf{$DB_{kwd}$}                     & 19,057                               & 50.5                      & 0.1                            & 37.4                                                                                   & 10,657.5                      & 5.1                  & 301.7                \\
\textbf{$DB_{centris}$}                 & 10,288                               & 50.5                      & 115.1                          & 20.0                                                                                   & 1,508,924.7                   & 4.4                  & -                    \\
\bottomrule[1.5pt]

\end{tabularx}
\label{tab:result_maintain}
\end{table*}
\vspace{0.05cm}
\noindent
\textbf{Characteristic Comparison.}
We compared DB$_{iot}$ with DB$_{kwd}$ and the databases constructed by \tool{Centris} (i.e., DB$_{centris}$) and \tool{TPLite} (i.e., DB$_{tplite}$) from three aspects: \emph{storage consumption}, \emph{efficiency} and \emph{maintainability}.

The comparison results are listed in Table~\ref{tab:result_maintain}.

In total, 
    DB$_{iot}$ includes only 1,872 libraries related to IoT, requiring 3.5 GB of storage, but DB$_{kwd}$ and DB$_{centris}$ consume ten times more storage because they include many irrelevant libraries through random TPL selection. 
In contrast, 
    \tplname effectively filters out most irrelevant libraries when constructing DB${iot}$.

When processing each TPL, 
    although DB$_{iot}$, DB$_{kwd}$, and DB$_{centris}$ cost the same time to clone one TPL repository and calculate function hash values within it, 
    DB$_{centris}$ needs 115.1s on average to eliminate the redundant functions within each TPL whereas DB$_{iot}$ only needs 0.1s to remove the redundancy. As most detection tools adapt similarity comparisons to identify the reused TPLs, 
    the database accuracy significantly affects the search space and detection efficiency.
Specifically,
    the search space of DB$_{iot}$ is smaller than both DB$_{kwd}$ and DB$_{centris}$, indicating that when using DB$_{iot}$, much fewer similarity comparisons are needed to identify one TPL reuse.

With frequent updates to TPLs, 
    databases for TPL reuse detection need to be updated and maintained easily. 
However, 
    the time needed to update DB$_{centris}$ is more than 100 hours.

On the contrary, 
    DB$_{iot}$ requires only a few minutes to update, 
    supported by hashing-index based elimination to remove redundancy.
Although \tool{TPLite} also creates a database for TPL reuse detection, 
    its database creation requires large storage to store excessive library details, including all function details.
Its storage capacity became overwhelmed when only processing 1,000 TPLs,
    indicating that using \tool{TPLite} to build a database for TPLite is impractical.

\vspace{0.05cm}
\noindent
\textbf{Database Integration.}
To test how DB$_{iot}$ performs when integrate with state-of-the-art tools, 
    we compared DB$_{iot}$, DB$_{kwd}$ and DB$_{centris}$ in detecting TPL reuses on the Top-10 IoT projects in C/C++ from GitHub as the target programs. 
Details of these projects are listed in Appendix~\ref{appendix_d}.

Table~\ref{tab:TPL_reuse_Centris} shows the results when integrating \tool{Centris} with the three different databases.
When using DB$_{centris}$, \tool{Centris} only identified 71 TPL reuses;
    however, its performance improved significantly, 
    detecting 174 and 112 TPL reuses when leveraging DB$_{iot}$ and DB$_{kwd}$, respectively. Notably, although DB$_{iot}$ has the smallest size, its comprehensiveness and specification enabled \tool{Centris} to detect more reuses than the other two larger databases. 

\begin{table}
\centering
\caption{TPL Reuse Detection Result of Centris}
\begin{tabularx}{0.77\linewidth}{c|cc|cc|cc}
\toprule[1.5pt]
\multirow{2}{*}{\textbf{Target}} & \multicolumn{2}{c|}{\textbf{DB$_{centris}$}} & \multicolumn{2}{c|}{\textbf{DB$_{iot}$}} & \multicolumn{2}{c}{\textbf{DB$_{kwd}$}} \\  \cmidrule(lr){2-3} \cmidrule(lr){4-5} \cmidrule(lr){6-7} 
 & \multicolumn{1}{c}{\textbf{Dtc}} & \multicolumn{1}{c|}{\textbf{Cfm}}  & \multicolumn{1}{c}{\textbf{Dtc}} & \multicolumn{1}{c|}{\textbf{Cfm}}  & \multicolumn{1}{c}{\textbf{Dtc}} & \multicolumn{1}{c}{\textbf{Cfm}}  \\ \midrule
AliOS-Things & \multicolumn{1}{c}{22} & \multicolumn{1}{c|}{15}  & \multicolumn{1}{c}{81} & \multicolumn{1}{c|}{36}  & \multicolumn{1}{c}{78} & \multicolumn{1}{c}{21}  \\ 
LiteOS & \multicolumn{1}{c}{15} & \multicolumn{1}{c|}{6}  & \multicolumn{1}{c}{26} & \multicolumn{1}{c|}{9}  & \multicolumn{1}{c}{23} & \multicolumn{1}{c}{4}  \\ 
Tasmota & \multicolumn{1}{c}{28} & \multicolumn{1}{c|}{17} & \multicolumn{1}{c}{93} & \multicolumn{1}{c|}{55} & \multicolumn{1}{c}{85} & \multicolumn{1}{c}{41}  \\ 
TizenRT & \multicolumn{1}{c}{31} & \multicolumn{1}{c|}{22}  & \multicolumn{1}{c}{72} & \multicolumn{1}{c|}{37} & \multicolumn{1}{c}{78} & \multicolumn{1}{c}{22}  \\ 
kamailio & \multicolumn{1}{c}{3} & \multicolumn{1}{c|}{1}  & \multicolumn{1}{c}{9} & \multicolumn{1}{c|}{2}  & \multicolumn{1}{c}{11} & \multicolumn{1}{c}{2}  \\ 
mbed-os & \multicolumn{1}{c}{7} & \multicolumn{1}{c|}{3}  & \multicolumn{1}{c}{22} & \multicolumn{1}{c|}{10}  & \multicolumn{1}{c}{35} & \multicolumn{1}{c}{11}  \\ 
openthread & \multicolumn{1}{c}{2} & \multicolumn{1}{c|}{2}  & \multicolumn{1}{c}{5} & \multicolumn{1}{c|}{2}  & \multicolumn{1}{c}{7} & \multicolumn{1}{c}{1}  \\ 
Sming & \multicolumn{1}{c}{5} & \multicolumn{1}{c|}{3}  & \multicolumn{1}{c}{48} & \multicolumn{1}{c|}{18}  & \multicolumn{1}{c}{10} & \multicolumn{1}{c}{5}  \\ 
TDengine & \multicolumn{1}{c}{3} & \multicolumn{1}{c|}{1}  & \multicolumn{1}{c}{4} & \multicolumn{1}{c|}{3}  & \multicolumn{1}{c}{9} & \multicolumn{1}{c}{4}  \\ 
zephyr & \multicolumn{1}{c}{1} & \multicolumn{1}{c|}{1}  & \multicolumn{1}{c}{3} & \multicolumn{1}{c|}{2}  & \multicolumn{1}{c}{8} & \multicolumn{1}{c}{1}  \\ \midrule 
\textbf{Total} & \multicolumn{1}{c}{\textbf{117}} & \multicolumn{1}{c|}{\textbf{71}} & \multicolumn{1}{c}{\textbf{363}} & \multicolumn{1}{c|}{\textbf{174}}  & \multicolumn{1}{c}{\textbf{334}} & \multicolumn{1}{c}{\textbf{112}} \\ 
\bottomrule[1.5pt]
\end{tabularx}
\begin{tablenotes}
\item[a] Dtc: Reuses detected.
\item[b] Cfm: Reuses confirmed with manual check.
\end{tablenotes}
\label{tab:TPL_reuse_Centris}
\end{table}

\subsubsection{Vulnerability Segment Evaluation}

We compared \sysname against the state-of-the-art CVE patch mapping tool VFCFinder, 
    commonly employed schemes Link Matching (directly collect patch information from CVE/NVD references)~\cite{v1scan} and \tool{SNYK} Database Retrieval~\cite{Snyk_Vulnerability_Database}, assessing their mapping accuracy, 
    time cost, 
    and space cost, respectively.
The patch commits can be directly collected from \tool{CVE/NVD} and are assumed available for all the solutions.

We exclude \tool{PatchScout}~\cite{patchscout} and \tool{VCMatch}~\cite{vcmatch} from the comparison because both schemes demonstrably underperform compared to \tool{VFCFinder}~\cite{vfcfinder}, 
    as discussed and evaluated in the study.


\vspace{0.05cm}
\noindent
\textbf{Commit benchmark.}
We randomly selected 200 CVEs and manually identified the corresponding patch commits using information from the NVD website and project commits. 
Overall, 
    158 (out of 200) CVEs have been labeled with the corresponding patch commits and the other 42 (out of 200) CVEs are labeled with ``None'' whose patch commits cannot be found within the given CPE scope manually (from the last vulnerable version to the fixed version).



\vspace{0.05cm}
\noindent
\textbf{Mapping accuracy.} 
We quantified the number of patch commits identified by each of the three methods and manually validated the 
    results. 
Table~\ref{table:patch_lcoal} presents the number of patch commits identified by each method (\textbf{Detected}) and the number 
    of correctly identified patches confirmed through manual verification (\textbf{Confirmed}). 
Additionally, 
    to quantify the overall performance of each method, we also considered their ability to handle cases labeled with ``None" and calculated the F1 score (\textbf{F1}).
As shown in the table, 
    \sysname successfully identified the highest number of validated patch commits (124) while generating \textbf{zero} false alarms.
For the 34 patch commits that were not identified, 
    we found that 33 were due to the absence of any information related to the vulnerable elements (files, functions, variables) in the CVE descriptions for revealing vulnerability details.
For instance, 
    the description of CVE-2007-0457~\footnote{https://nvd.nist.gov/vuln/detail/CVE-2007-0457} in the NVD is ``\textit{Unspecified vulnerability in the IEEE 802.11 dissector ..."}, which provides no useful information about the vulnerable elements. 
The one remaining failed case occurred because the LLM (\tool{GPT-4}) was unable to accurately identify the patch commit among the candidates in Step 4 (\ref{llmmss:step3}). Instances of this kind of failure are relatively rare.
In general,
    in terms of the F1 score, 
    \sysname achieved the highest F1 score among all approaches with a significant advantage.

To prevent data bias and ensure even data distribution,  
    we divided the dataset into five groups and performed 5-fold group validation to verify the performance of \sysname. 
\sysname achieved F1 scores of 87.5\%, 88.9\%, 88.51\%, 88.50\%, and 86.11\% in each validation round, respectively, demonstrating its effectiveness.

\begin{table}[t]
\centering
\caption{Performance of Security Patch Mapping}
\label{table:patch_lcoal}
\begin{tabularx}{0.8\linewidth}{c|c|c|c}
\toprule[1.5pt]
\textbf{Scheme}         & \textbf{Detected} & \textbf{Confirmed} & \textbf{F1 (\%)} \\
\midrule
Link Matching           & 47           & 47             & 45.86            \\
Snyk Database Retrieval & 60           & 60             & 55.05            \\
VFCFinder (Top-1)       & 158          & 79             & 61.69            \\
VFCFinder (Top-5)       & 158          & 98             & 71.53            \\
\sysname                 & 124          & 124            & 87.94            \\
\bottomrule[1.5pt]
\end{tabularx}
\end{table}

\begin{table}[t]
\centering
\caption{Time and Space Cost of Security Patch Mapping}
\label{table:time_space_cost}
\begin{tabularx}{0.7\linewidth}{c|c c|c}
\toprule[1.5pt]
\multirow{2}{*}{\textbf{Scheme}} & \multicolumn{2}{c|}{\textbf{Time Cost (s)}}          & \textbf{Space Cost (MB)} \\  \cmidrule(lr){2-3}  
                                 & \multicolumn{1}{c}{\textbf{Mean}} & \textbf{Median} & \textbf{Mean}            \\ \midrule
VFCFinder                        & \multicolumn{1}{c}{285.92}        &    71.90       & 337.94                   \\ 
\sysname                           & \multicolumn{1}{c}{84.68}         & 42.31           & 0.00                        \\
\bottomrule[1.5pt]
\end{tabularx}
\end{table}

\vspace{0.05cm}
\noindent
\textbf{Time and space cost.}
We compared the time and space overhead of \sysname and \tool{VFCFinder}. 
To evaluate the time overhead performance, we randomly selected 50 CVEs that all the two methods could successfully identify the patch commits. 
For space overhead, 
    we calculated the average space expenditure required to match each CVE directly on the ground truth dataset. 
The experimental results are shown in Table \ref{table:time_space_cost}.
As indicated, 
    \sysname exhibited the lowest time overhead 
    (in both average and median values) and the smallest space overhead. 
In terms of space overhead, 
    \tool{VFCFinder} requires cloning the GitHub repository locally to obtain candidate commits, 
    which results in significant additional space overhead.
In contrast, 
    \sysname accesses candidate commits via the GitHub API and only needs to use commits at runtime as temporary variables, 
    incurring no additional space overhead.

\vspace{0.05cm}
\noindent
\textbf{The Impact of Search Space on Time Cost.}
The time cost required for identifying patch commits is directly related to the number of candidate commits due to the difference in search space.
We developed an empirical study on the 158 CVEs which are labeled with patch commits in the ground truth dataset.
    This study aimed to reveal how the time cost of different methods varies with the distribution of the number of candidate commits.
    
\begin{figure}[t]  
  \centering
  \includegraphics[width=1\linewidth]{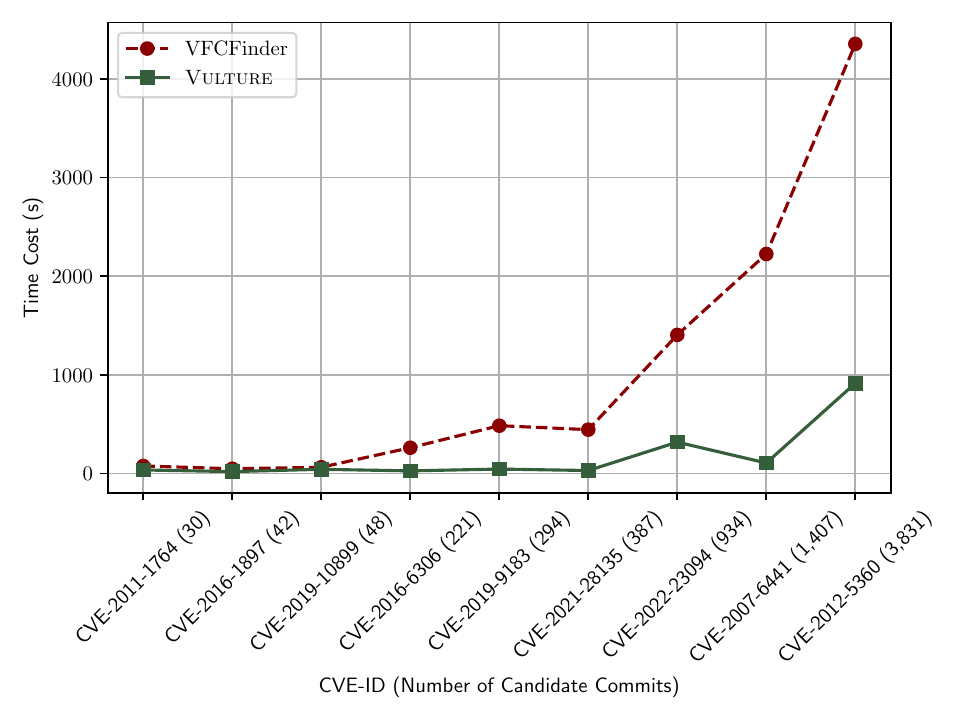}
  \caption{Correlation between the number of candidate commits and the time cost}
  \label{fig:space_time_correlation}
\end{figure}

Here, we categorized the distribution of candidate commit numbers \textit{N} into three ranges: 
    \ding{182} $100 > N >0$, 
    \ding{183} $1000 > N \geq 100$, 
    and 
    \ding{184} $N \geq 1000$. 
Among the total 158 labeled CVEs, the distribution of the number of commits  is as follows: \ding{182} 96, \ding{183} 54, and \ding{184} 8.
Furthermore, 
    we selected a total of 9 CVEs covering the three candidate commit number ranges
    to analyze the time cost of different approaches and evaluate their usability. 
As shown in Figure~\ref{fig:space_time_correlation},
    the horizontal axis represents the CVE ID with the number of candidate commits in the brackets; 
the vertical axis denotes the time overhead, measured in seconds.
As observed, 
    the time overhead of each approach positively correlates with the number of candidate commits.
Besides, 
    with the number of candidate commits increases, 
    the time overhead of \tool{VFCFinder} becomes substantially large;
in contrast, 
    even when the number of candidate commits reaches thousands, 
    \sysname can still keep the time overhead within a few hundred seconds.

\vspace{0.05cm}
\noindent
\textbf{Security Patch Mapping for CVE Database.}
\label{scrutnize1}

\subsection{Benchmark Vulnerability Detection}
\label{subsec:vul_detect}
We further assessed how \sysname performs when detecting 1-day vulnerabilities introduced by the reused TPLs.

We compared \sysname with the state-of-the-art work, \tool{V1SCAN}. 
The other tools were excluded because of various constraints. 
In detail, \tool{OSSFP} is not available as open-source, and \tool{SNYK} requires information on copyright and license, which is incompatible with our benchmark.


\vspace{0.05cm}
\noindent
\textbf{Vulnerability benchmark.}
To the best of our knowledge, there is no existing benchmark specifically for 1-day vulnerabilities in C/C++ programs. Therefore, we manually constructed a ground truth vulnerability benchmark. The benchmark comprises 200 vulnerable function reuse cases sourced from various repositories with more than 100 stars on GitHub. These 200 cases cover the reuse of 66 CVEs. Among these cases, 45\% have patched the CVE they reused, while 55\% have not. Additionally, 64\% of these cases involve custom reuse, including function name modification, statement modification, and operation modification.


\begin{table}[]
\centering
\caption{Vulnerability Detection Result on Ground Truth}
\begin{tabular}{c|c|cc|c}
\toprule[1.5pt]
\multirow{2}{*}{\textbf{Scheme}}  & \multirow{2}{*}{\textbf{Item}} & \multicolumn{2}{c|}{\textbf{Reuse Type}} & \multirow{2}{*}{\textbf{Total}} \\  \cmidrule{3-4}
                         &                       & \textbf{Custom Reuse}    & \textbf{Exact Reuse}   &                        \\ \midrule
\multirow{4}{*}{\textbf{\sysname}} & \textbf{Dtc-P}              & 72              & 19            & 91                     \\
                         & \textbf{Cfm-P}              & 65              & 19            & 84                     \\
                         & \textbf{Dtc-N}              & 87              & 22            & 109                    \\
                         & \textbf{Cfm-N}              & 78              & 22            & 100                    \\ \midrule
\multirow{4}{*}{\textbf{V1SCAN}}  & \textbf{Dtc-P}              & 51              & 10            & 61                     \\
                         & \textbf{Cfm-P}              & 35              & 10            & 45                     \\
                         & \textbf{Dtc-N}              & 66              & 14            & 80                     \\
                         & \textbf{Cfm-N}              & 42              & 13            & 55                     \\ \bottomrule[1.5pt]
\end{tabular}
\begin{tablenotes}
\item[a] Dtc: Vulnerabilities been detected.
\item[b] Cfm: Vulnerabilities been confirmed with manual check.
\item[c] P: Results on patched vulnerabilities.
\item[d] N: Results on non-patched vulnerabilities.
\end{tablenotes}
\label{tab:ground truth}
\end{table}
\vspace{0.05cm}
\noindent
\textbf{Scrutinize of results.}
The results are shown in Table ~\ref{tab:ground truth}. We compared \sysname with \tool{V1SCAN}. For custom reuse, \sysname successfully identified 65 patched and 78 unpatched cases, with only four false negatives and 12 false positives. In contrast, \tool{V1SCAN} identified 35 patched and 42 unpatched cases. For exact reuse, \sysname identified 19 patched and 22 unpatched cases without any false negatives or false positives, whereas \tool{V1SCAN} detected 10 patched and 13 unpatched cases with one false positive. Overall, \sysname achieved a 95.8\% F1 score, significantly outperforming \tool{V1SCAN}, which achieved only a 66.7\% F1 score.


After manually being confirmed by three PhD students from computer science majors, we observed that \tool{V1SCAN} produced a significant number of false negatives due to the limited comprehensiveness of its database. Additionally, it generated many false positives, which stemmed from inherent design flaws in its detection algorithm, as detailed in two motivational examples and Appendix \ref{appendix_c}. \sysname reported 4 false negatives and 12 false positives in total. Out of these, 4 false negatives and 8 false positives were a result of the limitations inherent in chunk-based detection methods. Chunk-based method allows us to pinpoint modified code lines only within their respective control statements. If a modified line is not associated with any control statement, its precise location remains unidentifiable, which can result in inaccuracies. 
\label{heavy _modification}
Additionally, the other 4 false positives arose from the limitations of similarity-based detection. If the reused function is significantly modified, the $Hash(i)$ and $H(j)$ can't be identified accurately (\ref{subsec:candidate_detection}), leading \sysname to miss recognizing the actual reused function. Nevertheless, these issues are relatively infrequent in real-world scenarios.

\subsection{Reuse and Vulnerability Detection In the Wild}
To ensure the robustness of our evaluation, we tested how does \sysname performs when facing real-world software. The target programs selected for this analysis are consistent with those listed in Table \ref{tab:TPL_reuse_Centris}. We then conducted a comprehensive analysis involving both TPL reuse detection and 1-day vulnerability detection using \sysname. 

\subsubsection{TPL reuse detection}
\begin{table}
\caption{TPL Reuse Detection Result of \sysname}
\centering
\begin{tabularx}{0.8\linewidth}{c|ccc|ccc}
\toprule[1.5pt]
\multirow{2}{*}{\textbf{Target}} & \multicolumn{3}{c|}{\textbf{DB$_{iot}$}}                                                                   & \multicolumn{3}{c}{\textbf{DB$_{kwd}$}}                                                                   \\ \cmidrule(lr){2-4} \cmidrule(lr){5-7}
                                 & \multicolumn{1}{c}{\textbf{Dtc}} & \multicolumn{1}{c}{\textbf{Cfm}} & \textbf{P} & \multicolumn{1}{c}{\textbf{Dtc}} & \multicolumn{1}{c}{\textbf{Cfm}} & \textbf{P} \\ \midrule
AliOS-Things                     & \multicolumn{1}{c}{47}                & \multicolumn{1}{c}{33}                 & 0.70               & \multicolumn{1}{c}{61}                & \multicolumn{1}{c}{20}                 & 0.33               \\ 
LiteOS                           & \multicolumn{1}{c}{18}                & \multicolumn{1}{c}{12}                 & 0.66               & \multicolumn{1}{c}{19}                & \multicolumn{1}{c}{4}                  & 0.21               \\ 
Tasmota                          & \multicolumn{1}{c}{66}                & \multicolumn{1}{c}{54}                 & 0.82               & \multicolumn{1}{c}{63}                & \multicolumn{1}{c}{39}                 & 0.62               \\ 
TizenRT                          & \multicolumn{1}{c}{44}                & \multicolumn{1}{c}{34}                 & 0.77               & \multicolumn{1}{c}{51}                & \multicolumn{1}{c}{20}                 & 0.39               \\ 
kamailio                         & \multicolumn{1}{c}{7}                 & \multicolumn{1}{c}{2}                  & 0.29               & \multicolumn{1}{c}{10}                & \multicolumn{1}{c}{2}                  & 0.20               \\ 
mbed-os                          & \multicolumn{1}{c}{17}                & \multicolumn{1}{c}{12}                 & 0.70               & \multicolumn{1}{c}{28}                & \multicolumn{1}{c}{10}                 & 0.36               \\ 
openthread                       & \multicolumn{1}{c}{4}                 & \multicolumn{1}{c}{2}                  & 0.50               & \multicolumn{1}{c}{4}                 & \multicolumn{1}{c}{1}                  & 0.25               \\ 
Sming                            & \multicolumn{1}{c}{32}                & \multicolumn{1}{c}{24}                 & 0.75               & \multicolumn{1}{c}{3}                 & \multicolumn{1}{c}{3}                  & 1.00               \\ 
TDengine                         & \multicolumn{1}{c}{3}                 & \multicolumn{1}{c}{3}                  & 1.00               & \multicolumn{1}{c}{4}                 & \multicolumn{1}{c}{2}                  & 0.50               \\ 
zephyr                           & \multicolumn{1}{c}{2}                 & \multicolumn{1}{c}{2}                  & 1.00               & \multicolumn{1}{c}{5}                 & \multicolumn{1}{c}{1}                  & 0.20               \\ \midrule
\textbf{Total}                            & \multicolumn{1}{c}{\textbf{240}}               & \multicolumn{1}{c}{\textbf{178}}                & \textbf{0.74 }              & \multicolumn{1}{c}{\textbf{248}}               & \multicolumn{1}{c}{\textbf{102}}                & \textbf{0.41 }              \\
\bottomrule[1.5pt]
\end{tabularx}
\begin{tablenotes}
\item[a] Dtc: Reuses detected.
\item[b] Cfm: Reuses confirmed with manual check.
\end{tablenotes}
\label{tab:TPL_reuse_Vulture}
\end{table}

In this experiment, we executed \sysname to detect the reused TPLs in the target programs by utilizing DB$_{iot}$ and DB$_{kwd}$, respectively.

Table~\ref{tab:TPL_reuse_Vulture} demonstrates the detection results.
\sysname correctly detected 178 reused TPLs when using DB$_{iot}$, achieving a precision of 74\%; however its precision is only 41\% when integrating DB$_{kwd}$.
Upon manually inspecting the failed cases, we found that 18 out of 62 were caused by the non-library projects in the DB$_{iot}$. These projects, such as \tool{LuatOS}, are not libraries themselves but are still reused by many other software, making them difficult to filter out manually. These non-library projects affect the specification of the database and cause \sysname to mistakenly identify them as the reused TPL instead of the actual parent libraries. 
Nine failed cases are caused by 9 TPLs which are filtered out by \sysname because these libraries are not prevalent on \tool{GitHub} with a few stars (\(<100\) stars). 
Due to the preset thresholds, $TH_{hash}$ and $TH_{sim}$, 29 cases failed as a result of overestimating similarity. Unfortunately, adjusting the thresholds could potentially result in additional failed or missed cases. The final six failed cases were due to the reuse of prevalent functions by many TPLs, which prevented \sysname from accurately identifying the actual TPL being reused. Such prevalent functions can be categorized into cryptographic functions 
(e.g., \func{md5Update}, \func{parse\_hex4}), utility functions (e.g., \func{hammingDistance}), and common functions (e.g., \func{strcpy}).

The disparity in the detection results between using DB${_{iot}}$ and DB${_{kwd}}$ arises because DB${_{kwd}}$ contains many non-IoT-specific libraries.
This lack of specificity results in incorrect identification of parent TPL reuse as child TPL reuse, leading to a high number of false positives. In contrast, DB${_{iot}}$, as a curated subset focused on IoT-related libraries, minimizes such misidentifications and achieves higher specificity. 
We did not incorporate \sysname with DB$_{centris}$ due to its lack of critical information (e.g., birth time, file path information) required by \sysname.

\subsubsection{1-day vulnerability detection}
\begin{table}[t]
\centering
\caption{Vulnerability Detection Result in Wild Software}
\begin{tabularx}{0.8\linewidth}{c|cc|cc|cc}
\toprule[1.5pt]
\multirow{2}{*}{\textbf{Target}} & \multicolumn{2}{c|}{\textbf{\sysname}} & \multicolumn{2}{c|}{\textbf{SNYK}} & \multicolumn{2}{c}{\textbf{V1SCAN}} \\ \cmidrule(lr){2-3} \cmidrule(lr){4-5} \cmidrule(lr){6-7} 
 & \multicolumn{1}{c}{\textbf{Dtc}} & \textbf{Cfm} & \multicolumn{1}{c}{\textbf{Dtc}} & \textbf{Confm} & \multicolumn{1}{c}{\textbf{Dtc}} & \textbf{Cfm} \\ \midrule
AliOS-Things & \multicolumn{1}{c}{93} & 89 & \multicolumn{1}{c}{105} & 84 & \multicolumn{1}{c}{8} & 2 \\ 
LiteOS & \multicolumn{1}{c}{19} & 19 & \multicolumn{1}{c}{22} & 16 & \multicolumn{1}{c}{3} & 3 \\ 
TizenRT & \multicolumn{1}{c}{68} & 66 & \multicolumn{1}{c}{16} & 10 & \multicolumn{1}{c}{11} & 8 \\ 
Tasmota & \multicolumn{1}{c}{1} & 1 & \multicolumn{1}{c}{2} & 0 & \multicolumn{1}{c}{0} & 0 \\ 
TDengine & \multicolumn{1}{c}{0} & 0 & \multicolumn{1}{c}{3} & 1 & \multicolumn{1}{c}{0} & 0 \\ \midrule
\textbf{Total} & \multicolumn{1}{c}{\textbf{181}} & \textbf{175} & \multicolumn{1}{c}{\textbf{148}} & \textbf{111} & \multicolumn{1}{c}{\textbf{22}} & \textbf{13} \\
\bottomrule[1.5pt]
\end{tabularx}
\label{tab:wildresult}
\begin{tablenotes}
\item[a] Dtc: Vulnerabilities been detected.
\item[b] Cfm: Vulnerabilities been confirmed with manual check.
\end{tablenotes}
\end{table}
\begin{table}[h!]
\centering
\caption{Time cost of TPL reuse and 1-day vulnerability detection across different tools (in seconds)}
\begin{tabular}{c|c|c|c|c}
\toprule[1.5pt]
\multirow{2}{*}{\centering \textbf{Target}} & \multicolumn{2}{c|}{\textbf{VULTURE}} & \multicolumn{2}{c}{\textbf{V1SCAN}} \\ \cmidrule{2-5}
                        & \textbf{TPL reuse} & \textbf{1-day} & \textbf{TPL reuse} & \textbf{1-day} \\ \midrule
AliOS-Things            & 20.1              & 3.0                      & 23.5               & 11.8                     \\ 
LiteOS                  & 14.1               & 3.9                      & 28.1               & 8.7                      \\ 
Tasmota                 & 5.5                & 129.9                    & 7.1                & -                        \\ 
TizenRT                 & 9.2                & 2.1                      & 8.1                & 8.8                      \\ 
TDengine                & 37.2               & -                        & 59.4               & -                        \\ 
\bottomrule[1.5pt]
\end{tabular}
\label{tab:tpl_v1scan_time}
\begin{tablenotes}
\footnotesize
\item \textit{The values in the table represent the average time (in seconds) required to detect a single TPL reuse or a single 1-day vulnerability. A dash ("-") indicates that no reuses or vulnerabilities were identified.}
\end{tablenotes}
\end{table}

Here, we applied \sysname, \tool{V1SCAN} and \tool{SNYK} to identify real-world 1-day vulnerabilities. Table \ref{tab:wildresult} only highlights the top 5 target programs with the highest number of detected vulnerabilities. The remaining 5 target programs were found to have no vulnerabilities according to the detection results.

Table \ref{tab:wildresult} shows that \sysname is the most effective tool, identifying 175 vulnerabilities,
    and consistently outperforms \tool{SNYK} and \tool{V1SCAN} across all test targets.

In comparison,
    \tool{SNYK} detected 111 vulnerabilities, and \tool{V1SCAN} detected only 13.
Furthermore,
    \sysname also pinpointed the exact locations and specific code statements that required to be patched, 
    and provided 154 GitHub patch commit URLs for the identified vulnerabilities.
However, 
    \tool{SNYK} only reported the CVE ID and provided 59 GitHub patch commit URLs, without offering further detailed information about the specific code requiring patching, and \tool{V1SCAN} provided only 13 GitHub patch commits and limited information about vulnerabilities, such as the vulnerable file and function names,

While manually verifying the missed cases, 
    we observed that \tool{SNYK} heavily relies on version information maintained by the target program. 
When a program does not maintain such information well, such as TizenRT, 
    \tool{SNYK} will miss numerous vulnerabilities.
When analyzing 87 vulnerabilities missed by \tool{V1SCAN}, 
    we found that 64 were caused by its limited database and others resulted from its coarse-grained line-matching approach.
Besides, 
    \tool{SNYK} and \tool{V1SCAN} incorrectly reported 37 and 9 vulnerabilities, respectively. 
Our manual checks revealed that these inaccuracies stemmed from \tool{SNYK}'s reliance on version-based detection and \tool{V1SCAN's} coarse-grained line-matching methods.




While analyzing the six vulnerabilities missed by \sysname, but detected by \tool{SNYK} and \tool{V1SCAN}, 
    three of them were missed due to the absence of patch commits which are mandatory for \sysname to locate patches. 
The remaining cases involved extensive code modifications, 
    which caused \sysname to fail in pairing the reused functions.



\subsubsection{Time cost} 

We assessed the time cost of \sysname and \tool{V1SCAN} in detecting TPL reuse and 1-day vulnerabilities.
Table~\ref{tab:tpl_v1scan_time} illustrates the time cost. 
Typically,
    \sysname costs no more than 25s to recognize all TPL reuses and less than 5s to identify a 1-day vulnerability even when handling large programs like \tool{AliOS-Things}.
In contrast.
    \tool{V1SCAN} usually takes more than 25s to locate a reuse and more than 5s to report a 1-day vulnerability.
When analyzing programs with extensive TPL reuses (e.g., \tool{Tasmota}), 
    \sysname takes longer to identify a single 1-day vulnerability, 
    but \tool{V1SCAN} times out without reporting any.



\subsection{Limitation}

\vspace{0.05cm}
\noindent
\textbf{Database Accuracy.} 
The performance of \sysname heavily relies on the database quality.
Since \sysname uses vulnerable elements and patched code to identify 1-day vulnerabilities, 
    poor quality or inconsistent formats in the data provided by NVD significantly impact the performance of \sysname.
Furthermore, 
    \sysname utilizes LLM to capture vulnerable elements for patch identification,
    which makes \sysname highly dependent on the performance of LLM.
Although LLM performs well generally when processing natural language contexts, 
   manual verification might still be necessary to ensure the accuracy of patch mapping. 
Otherwise, the performance of the vulnerability detection can be negatively affected. 


\vspace{0.05cm}
\noindent
\textbf{Similarity Comparison Limitation.} 
Although \sysname avoids using function or line matching to identify custom reuses, 
    the employed LSH comparison can still bring inaccuracies when processing custom reuses that involve extensive modifications.
The settings of similarity thresholds $TH_{hash}$ need to be adjusted to match various databases and accommodate different application scenarios.


\vspace{0.05cm}
\noindent
\textbf{Dataflow Restrictions in Chunk-based Analysis.} 
As \sysname generates chunks by analyzing code diff and the involved variables,
    it may cause incomplete data flows within chunks, 
    leading to certain information loss.
Consequently,
    \sysname performs poorly when analyzing target programs with extensive code modifications. 
In addition, 
    the accuracy of chunk generation also depends on the recognition results of vulnerable elements extracted from each CVE description.
Correspondingly, 
    chunk-based analysis is also affected the capability of LLM in processing natural language descriptions.


\section{Related Work}
\label{sec:related}

Sections A, B, and C present the related work on \textit{TPL Reuse Detection}, \textit{Security Patch Collection}, and \textit{TPL Vulnerability Detection}, respectively. 

\subsection{TPL Reuse Detection}


TPL reuse detection is an essential task aimed at identifying the TPLs that software relies on, 
    thereby facilitating comprehensive software maintenance and management~\cite{lopes2017dejavu}.
Several detection approaches 
    \cite{ma2016libradar, backes2016reliable, libD, Firmsec, ilibscope, SourcererCC, CCAligner, Centris, OSSFP, TPLite} targeting distinct platforms have been proposed.

TPL reuse tends to introduce similar or even identical code segments from TPL into software, 
    based on this,
    some studies proposed keyword token-based code reuse detection. 
SourcererCC~\cite{SourcererCC} leveraged token-based detection,
    which matches similar blocks using a bag-of-tokens-based strategy.
Further, 
    CCAligner~\cite{CCAligner} introduced the concept of a code window and additionally considered edit distance to detect large-gap code reuse. 
However, 
    token-based methods possessed poor performance in detecting customized TPL reuse, 
    and also cannot handle the issue of nested TPLs that interactive code segments.
Targeting the above issues, 
    function-level code reuse detection schemes are developed.
Woo et al.~\cite{Centris} proposed Centris to match unique parts of TPLs by hashing functions and eliminate duplicated code to extract function features. 
Similarly, 
    Wu et al.~\cite{OSSFP} developed OSSFP which focuses on identifying core functions. 
However, 
    Jiang et al.~\cite{TPLite} pointed out that in practical deployment, 
    Centris still exhibits poor performance in cases of TPL nesting. 
To address this, 
    TPLite~\cite{TPLite} was developed which introduced function birth time and the directory they locate in to build a dependency graph and analyze the nesting relationships.

Nevertheless, 
    the aforementioned methods require prior analysis of dependencies between TPLs before detection,
    which means the candidate TPL database cannot be easily expanded.
Adding any new TPL necessitates reanalyzing all dependencies — 
    an extremely time-consuming process.
Therefore, 
as TPL pool and software complexity increasingly develop, 
    TPL reuse detection tools must be maintainable, 
    which is compatible in \sysname.

\subsection{Security Patch Collection}

To collect security patches, 
    existing solutions include retrieving patch information from vulnerability maintenance platforms~\cite{cvefixes, v1scan, tracer} and mapping GitHub commits related to vulnerabilities from GitHub repositories~\cite{patchscout, vcmatch, patchmatch, vfcfinder}.
Some studies~\cite{cvefixes, v1scan}, 
    retrieved patch links by accessing the CVE/NVD websites and extracting the 
    ``Hyperlink" field. 
However, 
    due to the untimely updates of CVE/NVD, 
    the number of patches collected through this method is quite limited. 
Distinctively, 
    Tracer~\cite{tracer} collected security patches from multiple known sources, such as Debian and Red Hat, together with CVE/NVD. 
However,
    the issue of untimely updates is prevalent across all the platforms. 

To address the aforementioned issues, 
    other works~\cite{patchscout, vcmatch, patchmatch, vfcfinder}, 
    pursued an effort to identify patches from GitHub commits. 
Tan et al.~\cite{patchscout} introduced PatchScout, 
    which analyzed the correlation between vulnerability information and GitHub commits through a ranking strategy.
Similarly,
     \cite{vcmatch, patchmatch, vfcfinder} also developed ranking-based methods, 
     incorporating machine learning and deep learning models 
     (e.g., XGBoost~\cite{xgboost}, CNN~\cite{cnn}, CodeBERT~\cite{codebert}) to extract features. 
However, 
    ranking-based solutions cannot precisely map patch commits, 
    as depicted in the studies, highly recall rates can only be achieved in Top-N (e.g., Top-5) scenarios, which necessitates substantial manual efforts for verification.
Moreover, 
    these schemes cannot cope with the absence of patch commits for some CVEs.
If the CPE provided on CVE/NVD is wrong or the patch is just released by other channels, 
    false alarms would be returned with recognizing commits possessing some degree of similarity.


Differently, 
    \sysname filters candidate commits based on affected code elements,
    ensuring the exclusion of non-patch commits;
beyond that, 
    \sysname leverages LLM to logically determine the actual patch commits, 
    effectively addressing the defects of existing solutions.

\subsection{TPL Vulnerability Detection}

TPL vulnerability detection primarily relies on TPL reuse detection, 
    associated with TPL vulnerability analysis.
ReDeBug~\cite{redebug} employed a token-based approach for detecting reused vulnerable code through token-level similarity comparison.
Similarly, 
    Kim et al.~\cite{vuddy} proposed a high-efficiency TPL vulnerability detection strategy VUDDY based on function comparison, 
    with constructing hash-based function fingerprints.
Zhang et al.~\cite{zhang2018precise} designed FIBER, leveraging subgraph matching to trace the propagation of patches.
However, 
    these methods struggle to detect vulnerabilities in cases of complex custom TPL reuse, 
    limiting their practicality in real-world software.
To address the above issues, Xiao et al.~\cite{mvp} proposed MVP to extract coarse-grained semantic information for identifying modified vulnerable code clones.
Woo et al.~\cite{movery} introduced Movery, which constructs function fingerprints towards core vulnerability features based on vulnerability-specific code and patch lines.
Beyond that, 
    V1SCAN~\cite{v1scan} further extracted patched code lines from vulnerability patches and, 
    when detecting TPL vulnerabilities, 
    further reduced false alarms by comparing patch lines. 
Nevertheless, 
    these methods still exhibit limitations, 
    resulting in false positives in complicated custom reuse. 
In detail, 
    Movery fails to deal with the reused vulnerable code that has been patched; 
while V1SCAN solely relies on counting patch lines, instead of patch contents, 
    which is nearly tantamount to random judgment when handling some real-world software.
Additionally, 
    some studies focus on the analysis of TPL vulnerabilities in IoT firmware~\cite{onebadapple, Firmsec}, mobile software~\cite{OSSPolice, ma2016libradar, zhan2021atvhunter,LibScan} and web applications~\cite{shi2024recurscan},
    offering solutions tailored to other platforms and various programming languages.

\sysname employs static analysis to convert code to chunks based on vulnerability features.
Compared to existing approaches,
    \sysname can detect defects introduced by custom TPL reuse and analyze whether custom patch exists.
Additionally, 
    \sysname's TPL database is maintainable, 
    allowing it to adapt to the evolving software supply chain environment,
    a capability that existing solutions lack.

\section{Conclusion}
\label{sec:conclusion}
This study reveals that achieving accurate software supply chain security detection requires a comprehensive toolchain. 
We introduced \sysname, 
    which employs a combination of database construction, 
    security patch collection, 
    TPL reuse detection, 
    and library vulnerability detection. 
Through the construction of a comprehensive database and advanced algorithms, 
    \sysname minimizes false alarms and improves detection efficiency.
Evaluation shows that \sysname significantly outperforms state-of-the-art tools,
    including commercial solutions. 
Despite some limitations, 
    \sysname proves to be effective in real-world scenarios, 
    offering developers a robust tool to mitigate risks associated with TPL reuses.

\bibliographystyle{Format/IEEEtranS}

\bibliography{reference}

\appendix

\section{Appendix}
\label{sec:appendix}

\subsection{Appendix A}
\label{appendix_a}

In this section, 
    we present the complete process used by \sysname to identify the patch commit for a given vulnerability, 
    using CVE-2013-4080 as an instance.

The CVE/NVD provides a description for CVE-2013-4080: 
    ``The dissect\_r3\_upstreamcommand\_queryconfig function in epan/dissectors/packet-assa\_r3.c in the Assa Abloy R3 dissector in Wireshark 1.8.x before 1.8.8 does not properly handle a zero-length item, 
    which allows remote attackers to cause a denial of service 
    (infinite loop, and CPU and memory consumption) via a crafted packet".
The mapping process is detailed as follows:

\noindent
\textbf{1) LLM-based description parsing.}

\sysname analyzes the CVE description to parse vulnerable elements. 
The parsing results are as follows: 
\begin{itemize}
    \item File: packet-assa\_r3.c
    \item Function: dissect\_r3\_upstreamcommand\_queryconfig
    \item Variable: None
\end{itemize}

\noindent
\textbf{2) Slice-based commit filtering.}

First, 
    \sysname performs date-specific commit slicing to pinpoint the date range which contains the patch commit.
According to CVE/NVD, 
    "wireshark-1.8.7" is the last vulnerable version and "wireshark-1.8.8" is the fixed version, 
    narrowing the time range to (2013-05-17T16:41:42Z, 2013-06-07T15:49:07Z). 
A total of 348 commits fall within this period.

Next, 
    \sysname divides these commits into 18 slices, each containing 20 commits (the last slice contains only 8 commits).
It analyzes code changes in each slice by calculating the code changes between the first and the last commit, 
    marking those with vulnerable elements as candidate slices.
Only the third slice meets this criterion.

\noindent
\textbf{3) Candidate commit selection.}

\sysname sequentially analyzes the 20 commits within the candidate slice,
    selecting the commit that genuinely modifies the vulnerable elements as the candidate commit. 
Only the commit with the hash 779d28d39039ada8970c910d8350fc2eb05cf00a is identified as the candidate commit in this step.

\noindent
\textbf{4) LLM-based patch commit mapping.}

By combining the CVE description with the candidate commit,
    \sysname employs feature engineering to invoke an LLM (GPT-4.0) for analyzing whether the candidate commit serves as the patch for the vulnerability. 
Finally,
    \sysname successfully identifies the patch commit (779d28d39039ada8970c910d8350fc2eb05cf00a) for CVE-2013-4080.

\vspace*{-\baselineskip} 

\subsection{Appendix B}
\label{appendix_B}

\begin{table}[h!]
\centering
\caption{Target programs details}
\begin{tabular}{l|c|c|c}
\toprule[1.5pt]
\multirow{2}{*}{\textbf{Target Program}} & \multirow{2}{*}{\textbf{Stars}} & \multirow{2}{*}{\textbf{Lines}} & \multirow{2}{*}{\textbf{Version}}\\
 &  &  \\ \midrule
AliOS-Things & 4.5K & 4.62M & a99f207\\ 
LiteOS & 4.8K & 1.17M & 2f8fdf9\\ 
Tasmota & 21.4K & 1.59M & 4aa2da3\\
TizenRT & 0.5K & 2.16M & db112db\\
kamailio & 2.1K & 1.16M & 999d0c6\\
mbed-os & 4.6K & 8.64M & c7ea9c1\\
openthread & 3.4K & 451K & 2408c89\\
Sming & 1.4K & 32.33M & 895535e\\
TDengine & 22.8K & 774K & 8703373\\
zephyr & 9.7K & 2.18M & 2314a2c\\
\bottomrule[1.5pt]
\end{tabular}
\label{tab:target_program_detail}
\end{table}

The details of targeted projects used in the simulation are listed in Table \ref{tab:target_program_detail}. 
The "Stars" column shows the number of stars on GitHub, the "Lines" column indicates the number of lines of C/C++ code, and the "Version" column contains the commit hash for each program at the time of access.


\end{document}